\documentclass[letterpaper,twocolumn,10pt]{article}
\usepackage{usenix}

 \usepackage{algorithmic}
 \usepackage{booktabs}
 \usepackage{colortbl}
 \usepackage{xcolor}
 \usepackage{textcomp}
 \usepackage{rotating}
 \usepackage{fontawesome5}
 \usepackage{enumitem}
 \usepackage{multirow}
 \usepackage{subcaption}
 \usepackage{microtype}
 \usepackage{url}
 \usepackage{calc}
 \usepackage{float}
 \usepackage{listings}
 \usepackage{tikz}
 \usepackage[ruled,linesnumbered,algosection]{algorithm2e}

\usepackage[most]{tcolorbox}
\tcbset{
    mybreakbox/.style={
    enhanced,
    breakable,           
    colback=gray!5,
    colframe=gray!40,
    boxrule=0.2pt,
    arc=3pt,             
    left=6pt, right=6pt,
    top=6pt, bottom=6pt,
  },
}

\usepackage{amssymb}
\usepackage{amsmath}

\newcommand{\eff}[3]{\mathcal{E}(#1, #2, #3)}
\newcommand{\weff}[2]{\mathcal{E}_{web}({#1},{#2})}
\newcommand{\att}[1]{A_{\scriptstyle \text{#1}}}

\newcommand{\pol}[1]{P_{\scriptstyle \text{#1}}}
\newcommand{\amod}[1]{T_{\scriptstyle \text{#1}}}

\newcommand{\demb}{D_{\scriptstyle \text{embed}}}

 \newcommand{\code}[1]{{\small\ttfamily\hyphenchar\font=`\- #1}}

 \lstdefinelanguage{Pseudocode}{
   keywords={for, each, if, else, allow, reject, then, end, in, not, and, Input},
   keywordstyle=\color{blue}\bfseries,
   comment=[l]{\#},
   commentstyle=\color{gray}\itshape,
   string=[b]",
   stringstyle=\color{red},
 }

\newcommand{\cir}[1]{\tikz[baseline]{%
    \node[anchor=base, draw, fill, text=white, font=\bf\footnotesize, circle, inner sep=0, minimum width=0.8em]{#1};}}

 \newcommand{\bfparagraph}[1]{\noindent\textbf{#1}\,}

 \newcommand{\lighttester}{HPP}
 \newcommand{\thoroughtester}{EPP}
 \newcommand{\LightTester}{Heuristic Policy Profiler}
 \newcommand{\ThoroughTester}{Exhaustive Policy Profiler}

 \usepackage{hyperref}

\begin{document}

\date{}

\title{\Large \bf Fetch Me If You Can: A Comprehensive Analysis of Fetch Metadata}

\author{
    {\rm Pedro Bernardo, Alex Ploner, Marco Squarcina, Mauro Tempesta, Matteo Maffei}\\
    TU Wien
} 

\maketitle

\begin{abstract}

Effective server-side defenses against cross-site attacks depend on understanding the context in which browser requests are generated. However, servers often lack the information needed to distinguish legitimate requests from potentially malicious cross-site interactions. Fetch Metadata, a security mechanism introduced in 2019, addresses this limitation by adding a set of HTTP headers to outgoing browser requests. These headers provide servers with information about a request's context and provenance, enabling them to enforce request isolation policies accordingly. However, the extent of Fetch Metadata's real-world adoption and its interaction with existing security mechanisms remain largely unexplored.

In this paper, we present the first large-scale measurement study of Fetch Metadata adoption. We analyze the top 1M websites in the Tranco list to detect deployments of Fetch Metadata isolation policies and identify custom policies used in the wild. We then examine how these policies address cross-site threats and assess their security coverage. Finally, we provide the first systematic comparison of Fetch Metadata with existing client-side security mechanisms, revealing security gaps that arise from differences between client-side and server-side enforcement.

\end{abstract}

\section{Introduction}
The Web platform is the primary target for attackers due to its ubiquity and central role in our daily lives. It supports a wide range of APIs~\cite{webapis} that compose with each other while striving to maintain backward compatibility, resulting in a large and evolving attack surface. Although the platform supports several security mechanisms, the path toward default secure behavior remains challenging~\cite{janc2020places}. As a result, the burden of securing applications often falls onto developers, which can lead to insecure websites exposing users to security and privacy risks. Most existing security mechanisms are designed to mitigate specific attack vectors. For example, the Content Security Policy (CSP) mainly targets script injection and framing attacks, while the \code{SameSite} cookie attribute protects against Cross-Site Request Forgery (CSRF). More recent protections like the Cross Origin Opener Policy (COOP) can help securing web applications against certain types of Cross-Site Leaks (XS-Leaks). To deploy these mechanisms effectively, developers must keep up with the latest threats and understand which mechanisms can effectively mitigate different attack classes. This race between protections and defenses highlights the need for broader, general-purpose security mechanisms that offer strong protections across multiple attack classes while preserving functionality and compatibility.

Fetch Metadata (FM)~\cite{west_20_fetch} is a browser feature that tries to address this gap: browsers attach metadata to outgoing requests, giving servers additional context like a request's origin-relation, destination, and initiator. This information, previously unavailable to servers, allows for the implementation of server-side security policies, mitigating a broad range of cross-origin attacks. However, Fetch Metadata is a relatively recent mechanism (2019), and its adoption and effectiveness have not been studied yet. A central challenge is that FM policies remain invisible to the browser, making their detection impossible by directly examining HTTP response headers or analyzing client-side code as done in prior work for client-side defenses like CSP and Subresource Integrity (SRI)~\cite{WebAlmanac}.
We identify three fundamental research questions:
\begin{itemize}
    \item \textbf{RQ1.} What attack classes do Fetch Metadata isolation policies mitigate, and how effectively?
    \item \textbf{RQ2.} How does Fetch Metadata compare against existing security mechanisms, especially regarding the policy enforcement location (server-side vs. client-side)?
    \item \textbf{RQ3.} How is the adoption of Fetch Metadata policies in the wild, considering server-side support and deployment by websites?
\end{itemize}

\bfparagraph{Contributions.} In this paper, we answer all three research questions through the following contributions:
\begin{itemize}
    \item We introduce a lightweight and intuitive formal framework for evaluating the effectiveness of Fetch Metadata policies against multiple classes of cross-site attacks, including Cross-Site Request Forgery (CSRF), Cross-Site Script Inclusion (XSSI), Clickjacking, and Cross-Site Leaks (XS-Leaks) (Sec.~\ref{chap:isolationPolicies}). Our analysis revealed gaps and inaccuracies in the leading community-maintained XS-Leaks reference~\cite{XSLeaks}. To address these issues, we propose several updates~\cite{issue1, issue2, issue3, issue4} regarding the protection coverage provided by Fetch Metadata \textbf{(RQ1)}.

    \item Using our framework, we conduct the first formal comparative analysis of Fetch Metadata and existing security mechanisms (Sec.~\ref{chap:fetchComparison}), including Origin and Referer headers, Content Security Policy, X-Frame-Options, SameSite cookies, Cross-Origin Opener Policy, Cross-Origin Resource Policy, and X-Content-Type-Options. We show how Fetch Metadata can complement, strengthen, or replace existing defenses, while also identifying inherent limitations of client-side enforcement. We empirically validate our findings and present proof-of-concept timing attacks against popular client-side defenses that are mitigated by Fetch Metadata policies \textbf{(RQ2)}.

    \item We systematically analyze support for Fetch Metadata policies across popular server-side frameworks, CMSes, and web servers, and develop a methodology to infer the presence of such policies on live websites (Sec.~\ref{chap:methodology}). Using this approach, we conduct the first large-scale measurement study of Fetch Metadata adoption across the top 1M websites in the Tranco list. Our analysis reveals that only 0.23\% of sites deploy Fetch Metadata policies (Sec.~\ref{chap:evaluation}) \textbf{(RQ3)}.
    
\end{itemize}

\noindent Our study shows that Fetch Metadata offers a powerful and broadly applicable server-side defense, yet remains largely overlooked.
By analyzing its security guarantees, comparing it with other protections, and measuring its adoption, we expose some root causes behind its limited deployment. More broadly, this study highlights the need for practical mechanisms that offer strong, general-purpose protection as the Web continues to evolve, and distills lessons learned to guide the design of future security features.

\section{Background on Fetch Metadata}
\label{chap:background}
In this section, we describe Fetch Metadata and the relevant HTTP headers. Then, we define the attacker model we will use throughout the rest of the paper and the attacks we deem in scope for our analysis.

\subsection{Fetch Metadata Headers}
Fetch Metadata is a security feature that was proposed by Google in 2019 to provide additional information about the context of outgoing HTTP requests through a set of HTTP headers~\cite{west_20_fetch} of the form \code{Sec-Fetch-*}, which are automatically attached by the browser and cannot be modified by JavaScript. Note that these headers are only attached to requests to \emph{potentially trustworthy}~\cite{west_secure} URLs, most notably requests over \code{https://} or \code{wss://}. The additional context enables servers to decide whether to allow or reject requests based on their origin position, initiator, and intended use. For instance, a web application endpoint may permit same-origin requests generated by a form submission but block requests initiated by an image tag from a cross-site source. Currently, all major browsers support Fetch Metadata headers~\cite{wptFetchMetadata}:

   \begin{center}
      \small
      \begingroup
      \renewcommand{\arraystretch}{1.5}
      \begin{tabular}{@{}lp{0.68\linewidth}@{}}
      \code{Sec-Fetch-Site} & Informs a server whether a request for a resource originates from the same origin, same site, or from a direct user navigation. \\
      \code{Sec-Fetch-Mode} & Specifies the mode of a request~\cite{httpReqMode}, enabling servers to distinguish, e.g., between requests to load resources and navigations. \\
      \code{Sec-Fetch-Dest} & Indicates the request's destination, i.e., where the response data will be used. For example, an image gallery service may only allow requests with \code{Sec-Fetch-Dest: image}. \\
      \code{Sec-Fetch-User} & Included only in requests triggered by user interaction, and its value is always the string \code{?1}. For example, the header is included when a user clicks a link to another page. \\
      \end{tabular}
      \endgroup
      \end{center}

\bfparagraph{Example.}
If an image tag tries to load a resource from the same origin, the headers of the request will be:
\begin{center}
\begin{minipage}{0.95\linewidth}
\begin{lstlisting}[numbers=none]
    ...
    Sec-Fetch-Site:   same-origin
    Sec-Fetch-Mode:   no-cors
    Sec-Fetch-Dest:   image
\end{lstlisting}
\end{minipage}
\end{center}

\subsection{Attacker Models}
\label{sub:attackerModel}
The main goal of Fetch Metadata is to provide servers with more context about incoming requests so they can filter them based on properties like origin or target element. Therefore, FM can only be used to reason about attack flows that can be distinguished through these properties and where the integrity of browser-provided headers is not compromised. We identify two standard attacker models in the literature that are relevant to our analysis: the standard web attacker~\cite{csrf_barth}, and the same-site (or related-domain) attacker~\cite{same_site_squarcina, originCookies}.
A web attacker controls an arbitrary number of hosts and can lure the victim into visiting an arbitrary website. The same-site attacker controls a domain considered same-site in relation to the victim website (e.g., as a result of a subdomain takeover~\cite{Dangling_DNS,Cloud_Strife}). Two domains are considered part of the same site (or related) when they share a common eTLD+1 (Effective Top-Level Domain), also known as a registrable domain. This relationship holds irrespective of the depth of the subdomain. The URLs \url{https://foo.bar.example.com} and \url{https://foo.example.com} are considered related as they share the common eTLD+1 \url{example.com}. However, this relationship does not apply to different GitHub pages, such as \url{https://foo.github.io} and \url{https://bar.github.io}, because the domain \url{github.io} is included in the Public Suffix List (PSL)~\cite{PSL}. This list, maintained by Mozilla, defines the eTLDs and is imported by browsers to determine site relationships, e.g., to define the scope of cookies~\cite{same_site_squarcina}. In summary, the victim can visit attacker-controlled malicious websites or sites that include resources hosted by the attacker (e.g., scripts, HTML pages, images). However, the attacker cannot intercept network requests and responses to other websites.

\subsection{Web Attacks}
\label{sub:attacks}

We now outline the primary attacks executed under the aforementioned attacker models. Drawing from industry standards like the OWASP Top 10 \cite{OWASP_TOP10} and MITRE's CWE \cite{MITRE_CWE}, we focus on the following cross-site web attacks.

\bfparagraph{Cross-Site Request Forgery (CSRF)}.
In a CSRF attack, the attacker manipulates the victim's browser to send an authenticated request to a target website, performing some sensitive actions on behalf of the victim~\cite{csrf_barth}. There are special cases of CSRF attacks that differ from this definition, be it in goal or authentication requirements (e.g., login CSRF). But, they all share the requirement for an attacker-forged cross-site request. Therefore, for the rest of the paper, we group these attacks in the CSRF category. Same-site attacks, such as the CORF attack variant described in~\cite{cookie_crumbles} are not discussed in the rest of the paper for sake of clarity.

\bfparagraph{Cross-Site Script Inclusion (XSSI).}
XSSI occurs when a victim visits a malicious website that performs authenticated requests to load resources from a vulnerable website as scripts to exfiltrate user information~\cite{XSSI}. This exfiltration abuses the fact that the Same-Origin Policy (SOP) permits loading cross-origin scripts through \code{<script>} tags. Since the scripts are executed in the loading page's context, user-dependent information stored in the requested resources can be interpreted as a script and leaked to the attacker. While CSRF attacks break integrity assumptions by performing authenticated actions in the victim's name, XSSI is used to break confidentiality, i.e., stealing user data provided by authenticated resource requests.

\bfparagraph{Cross-Site Leaks (XS-Leaks).}
XS-Leaks~\cite{XSLeaks, SoK, COSI} allow attackers to bypass the SOP and exfiltrate user data across the origin boundary by abusing web features. In a typical setting, the attacker first identifies a resource on the target website that depends on the user's state (\emph{state-dependent resource}). The attacker loads that resource using an \emph{inclusion method} into a context where exfiltration is possible through a \emph{leak technique}. For example, if the number of frames in a webpage depends on the user's authentication status, an attacker could lure the victim into visiting a malicious website that opens the target webpage in a new window via \code{window.open}. The attacker can then count the number of frames in the opened window with the \code{window.frames} API to infer the user's session state.

\bfparagraph{Clickjacking.}
In Clickjacking \cite{Clickjacking, OWASP_Clickjacking} attacks, the victim visits a malicious website and is tricked into interacting with elements from another website to trigger unintended actions. This is typically achieved by loading the target website in a transparent iframe and overlapping relevant UI elements with benign-looking elements in the attacker's website. Using this technique, the attacker can hijack user clicks on the attacker's page to perform sensitive actions on other websites.

\section{Formal Model of Fetch Metadata}
\label{sec:formal-model}

In this section, we introduce a lightweight formal framework to systematically evaluate Fetch Metadata. 

Let the Request Space $\mathcal{R}$ be the set of all concrete HTTP requests. Each request $r \in \mathcal{R}$ carries many components (e.g., headers, body), but only some are relevant to FM. We define the FM projection $\pi_{\scriptstyle \text{FM}}$, which maps a request to the four attributes that FM policies can observe:
\[
\pi_{\scriptstyle \text{FM}}(r) = (\sigma, \; \mu, \; \delta, \; m)
\]

\vspace{-15pt}
   \begin{center}
      \small
      \begingroup
      \renewcommand{\arraystretch}{1.5}
      \begin{tabular}{@{}lp{0.6\linewidth}@{}}
      $\sigma \in \{\text{same-origin}, \text{same-site}, \text{cross-site}, \text{none}\}$ & \code{Sec-Fetch-Site} \\
      \parbox[t]{0.62\linewidth}{%
       $\mu \in \{\text{navigate}, \text{no-cors}, \text{cors}, \text{same-origin},$\\%
       \hspace*{2em}$\text{websocket}\}$} & \code{Sec-Fetch-Mode} \\
      $\delta \in \{\text{document}, \text{iframe}, \text{script}, \text{image}, \dots\}$ & \code{Sec-Fetch-Dest} \\
      $m \in \{\text{GET}, \text{POST}, \text{PUT}, \dots\}$ & HTTP method. \\
      \end{tabular}
      \endgroup
  \end{center}

We denote the FM-visible request space by $\mathcal{F} = \pi_{\scriptstyle \text{FM}}(\mathcal {R})$.
Since FM policies can only observe these four attributes, we can restrict our reasoning to $\mathcal{F}$ without loss of generality. The remaining request properties (e.g., initiator and target origins, cookies, embedding ancestor chain) remain accessible to other security mechanisms, as we explore in Sec.~\ref{chap:fetchComparison}.

\bfparagraph{Browser-Emittable Set ($\mathcal{R}_B$).}
The Cartesian product of request variables produces combinations that standard-compliant browsers cannot generate. For example, a \code{POST} request cannot be destined to an image or script tag, and user-initiated requests ($\sigma = \text{none}$) are exclusively top-level \code{GET} navigations. 
We define $\mathcal{R}_B \subseteq \mathcal{R}$ as the \emph{browser-emittable set}, i.e., the subset of $\mathcal{R}$ that satisfies the constraints imposed by browsers. The complete constraint set is listed in Appendix~\ref{appendix:browser-constraints}. These constraints are derived from the Fetch, HTML and Fetch Metadata Specifications~\cite{whatwg_fetch, whatwg_html,west_20_fetch}. We define $\mathcal{F}_B = \pi_{\scriptstyle \text{FM}}(\mathcal{R}_B)$ as the FM-visible request space of $\mathcal{R}_B$.

\bfparagraph{Attacker Models.}
We formalize attacker capabilities as sets of constructible HTTP requests. An attacker model $T$ is defined by its constructible request set $\mathcal{C}(T) \subseteq \mathcal{F}$. The browser sets $\sigma$ based on the origins of the request initiator and the server, while attackers can control $\mu$, $\delta$, and $m$. We define two standard attacker models:
\begin{align}
\mathcal{C}(\amod{web}) &= \{ r \in \mathcal{F}_B \mid r.\sigma \in \{\text{cross-site}, \text{none}\} \} \label{eq:web-attacker} \\
\mathcal{C}(\amod{same}) &= \mathcal{C}(\amod{web}) \cup \{ r \in \mathcal{F}_B \mid r.\sigma = \text{same-site} \} 
\label{eq:same-site-attacker}
\end{align}

By definition, $\mathcal{C}(\amod{same})$ contains all requests in $\mathcal{C}(\amod{web})$ plus same-site requests.
We include $\sigma = \text{none}$ in $\mathcal{C}(\amod{web})$ to model user-initiated navigations (e.g., phishing), which is a capability of the web attacker~\cite{AkhaweBLMS10}.

To reason formally about FM policies, we must define some primitives for our framework:

\bfparagraph{Policies.}
A policy $P$ is defined as a predicate $P: \mathcal{F} \to \{\bot, \top\}$ that, given a request $r$, accepts it ($P(r) = \top$) or rejects it ($P(r) = \bot$) according to a set of conditions.

\bfparagraph{Attack Classes.}
An attack class $A \subseteq \mathcal{F}$ is defined as the set of request characteristics required for an attack to succeed. This set-based representation enables us to reason precisely about which attack flows are blocked by a given policy.

\bfparagraph{Effectiveness Metric.}
Given an attacker model $T$, attack class $A$, and policy $P$, we define the effectiveness metric as:
\[
\mathcal{E}(T, A, P) = \{ r \in \mathcal{C}(T) \cap A \mid P(r) = \top \}
\]
The result is the set of attack requests that succeed despite policy enforcement. 
We classify the protection provided by policy $P$ against attack $A$ under the attacker model $T$ as:
\begin{description}[leftmargin=0pt, labelindent=0pt, itemsep=-2pt]
    \item \emph{Complete}: $\mathcal{E} = \emptyset$ (no attack flows succeed)
    \item \emph{Partial}: $\emptyset \subset \mathcal{E} \subset (\mathcal{C}(T) \cap A)$ (some attack flows blocked)
    \item \emph{None}: $\mathcal{E} = \mathcal{C}(T) \cap A$ (all attack flows succeed)
\end{description}

\section{Policy Effectiveness Analysis}
\label{chap:isolationPolicies}

Having established the formal primitives of our framework, we now apply this methodology to evaluate the security properties of standard isolation policies against the landscape of cross-site attacks introduced in Sec.~\ref{chap:background}.

\subsection{Isolation Policies}
\label{sub:isolationPolicies}

The Fetch Metadata request headers provide contextual information that enables servers to build effective security measures. These isolation policies inspect the values of the \code{Sec-Fetch-*} headers and determine whether to allow the HTTP request to proceed or reject it with an \code{403 Forbidden} response~\cite{west_20_fetch}. To prevent disclosure of sensitive timing information, policies should reject requests before any data processing (e.g., authentication) takes place~\cite{weichselbaum_2020_protect}.

We present the four standard isolation policies collected from documentation and online resources~\cite{XSLeaks, weichselbaum_2020_protect, Appspot}, and their formal definition. Pseudocode is provided in Appendix~\ref{appendix:isolationPolicies}.

\bfparagraph{Navigation Isolation Policy} (NIP) blocks all cross-site navigational requests, i.e., requests that change the current browsing context's document such as those initiated by clicking a link or submitting a form.
\begin{align*}
\pol{NIP}(r) = \bot \iff r.\mu = \text{navigate} \land r.\sigma = \text{cross-site}
\end{align*}

\bfparagraph{Resource Isolation Policy} (RIP) restricts resource loading to same-site requests and direct navigations. It rejects all cross-site requests except GET/HEAD navigational requests. Resources intended for cross-site contexts (e.g., CORS) can be explicitly allowed~\cite{weichselbaum_2020_protect}. A case study implementing RIP across 500+ Google services found only 3\% of endpoints required exemptions~\cite{SoK}.
\begin{align*}
\pol{RIP}(r) = \bot \iff r.\sigma = \text{cross-site} \land  \\
(r.m \notin \{\text{GET}, \text{HEAD}\} \lor r.\mu \neq \text{navigate}) 
\end{align*}

\bfparagraph{Framing Isolation Policy} (FIP) rejects all navigational requests originating from embedded elements. A relaxed variant (rFIP) allows same-origin requests. $\demb$ = \{frame, iframe, object, embed\} represents the set of embedding destinations.
\begin{align*}
\pol{FIP}(r) = \bot \iff r.\mu = \text{navigate} \land r.\delta \in \demb
\end{align*}

\bfparagraph{Strict Isolation Policy} (SIP) rejects all cross-site requests.
\begin{align*}
\pol{SIP}(r) = \bot \iff r.\sigma = \text{cross-site}
\end{align*}
Since $\pol{NIP}$ rejects cross-site navigational requests and $\pol{RIP}$ rejects cross-site non-navigational requests, their disjunction covers all cross-site requests: $$\pol{SIP}(r) = \bot \iff \pol{NIP}(r) = \bot \lor \pol{RIP}(r) = \bot$$

\subsection{Policies vs. Attacks}
\label{sub:protectionAgainstAttacks}

The policies presented in Sec.~\ref{sub:isolationPolicies} use \emph{site} as the security boundary, restricting cross-site requests with the exception of the FIP, which ignores the \code{Sec-Fetch-Site} header. As a consequence, the RIP, NIP and SIP are trivially ineffective against attacks performed under $\amod{same}$, and the FIP cannot distinguish between same-site and cross-site attackers. These policies are, therefore, tailored to the standard web attacker $\amod{web}$. Within this model, we evaluate their effectiveness against the attacks presented in Sec.~\ref{sub:attacks}, which rely precisely on cross-site requests. For conciseness, we define $\mathcal{E}_{web}$ as the partial application of $\mathcal{E}$ where $T=\amod{web}$. Therefore, $\weff{A}{P}=\eff{\amod{web}}{A}{P}$ represents the effectiveness of policy $P$ against attack $A$ under the web attacker model.

\begin{table}[h]
\centering
\small
\caption{Protection offered by isolation policies against different attacks as computed by $\weff{A}{P}$.}
\begin{tabular*}{\columnwidth}{@{\extracolsep{\fill}}l|c|c|c|c}
\toprule
\textbf{Attack Classes}       & \textbf{NIP} & \textbf{RIP} & \textbf{FIP} & \textbf{SIP} \\ \midrule
CSRF                          & \faAdjust & \faCircle & \faAdjust & \faCircle \\
XSSI                          & \faCircle[regular] & \faCircle & \faCircle[regular] & \faCircle \\
XS-Leaks                      & \faAdjust & \faAdjust & \faAdjust & \faCircle \\
Clickjacking                  & \faCircle & \faCircle[regular] & \faCircle & \faCircle \\ \bottomrule
\end{tabular*}
\caption*{\faCircle~Complete protection \faAdjust~Partial mitigation \faCircle[regular]~No mitigation}
\label{tab:PoliciesVsAttacks}
\end{table}

\bfparagraph{Cross-Site Request Forgery (CSRF).}
In a CSRF attack, the attacker causes the victim's browser to send an authenticated state-changing request to the target website~\cite{csrf_barth}. CSRF attacks are also possible if web applications implement state-changing operations through safe methods like GET~\cite{StateOfSameSite}. We consider this scenario as out of scope for our analysis and assume that web applications respect HTTP semantics. We formalize the CSRF attack class as:
\begin{align*}
\att{CSRF} &= \{ r \in \mathcal{F} \mid r.\sigma = \text{cross-site} \,\, \land \\
&r.m \in \{\text{POST}, \text{PUT}, \text{DELETE}\} \}
\end{align*}
\textit{Isolation Policy Effectiveness:}
\begin{align*}
\weff{\att{CSRF}}{\pol{RIP}} &= \weff{\att{CSRF}}{\pol{SIP}} = \emptyset \\
\weff{\att{CSRF}}{\pol{NIP}} &= \{ r \mid r.\mu \neq \text{navigate} \land {} \\
    &\quad r.m \in \{\text{POST}, \text{PUT}, \text{DELETE}\} \} \\
\weff{\att{CSRF}}{\pol{FIP}} &= \{ r \mid r.\mu \neq \text{navigate} \lor {} \\
    &r.\delta \notin \demb \}
\end{align*}
SIP and RIP provide complete protection ($\mathcal{E} = \emptyset$). For all $r \in \att{CSRF}$, $r.\sigma = \text{cross-site}$ and $r.m \in \{\text{POST}, \text{PUT}, \text{DELETE}\}$, which implies $\pol{SIP}(r) = \bot$ and $\pol{RIP}(r) = \bot$. NIP and FIP provide partial mitigation (non-navigational and non-framed requests succeed).

\bfparagraph{Cross-Site Script Inclusion (XSSI).}
XSSI relies on authenticated cross-site GET requests to load resources from a target website as scripts in the attacker's page.
\begin{align*}
\att{XSSI} &= \{ r \in \mathcal{F} \mid r.\sigma = \text{cross-site} \,\, \land \\ 
&r.\mu = \text{no-cors} \land r.\delta = \text{script} \land r.m = \text{GET} \}
\end{align*}

\textit{Isolation Policy Effectiveness:}
\begin{align*}
\forall P \in \{\pol{SIP}, \pol{RIP}\}, \quad \weff{\att{XSSI}}{P} &= \emptyset \\
\forall P \in \{\pol{FIP}, \pol{NIP}\}, \quad \weff{\att{XSSI}}{P} &= \att{XSSI}
\end{align*}
SIP and RIP provide complete mitigation. For all $r \in \att{XSSI}$, $r.\sigma = \text{cross-site}$, which implies $\pol{SIP}(r) = \bot$ and $\pol{RIP}(r) = \bot$, thus $\weff{\att{XSSI}}{P} = \emptyset$. NIP and FIP offer no mitigation.

\bfparagraph{Cross-Site Leaks (XS-Leaks).}
The main components of an XS-Leak attack are an inclusion method that loads a state-dependent resource in a cross-site context and a leak technique that allows the attacker to exfiltrate state-dependent information from the included resource. Leak techniques are performed client-side and do not involve network requests, making them orthogonal to FM isolation policies~\cite{xsinator,SoK}. However, inclusion methods require cross-site interactions to load state-dependent resources, and are well defined and understood in the literature~\cite{XSLeaks, SoK, xsinator}.
While XS-Leaks could theoretically use non-\code{GET} methods, defining the XS-Leaks attack class with an unconstrained request method introduces significant overlap with the definition of CSRF attacks. Furthermore, real-world XS-Leaks overwhelmingly rely on \code{GET} requests for resource inclusion~\cite{XSLeaks}. Therefore, we define the XS-Leak attack class as follows:
\begin{align*}
\att{XSL} &= \{ r \in \mathcal{F} \mid r.\sigma = \text{cross-site} \land r.m = \text{GET} \}
\end{align*}
\textit{Isolation Policy Effectiveness:}
\begin{align*}
\weff{\att{XSL}}{\pol{FIP}} &= \{ r \in \att{XSL} \mid r.\delta \notin \demb \} \\
\weff{\att{XSL}}{\pol{NIP}} &= \{ r \in \att{XSL} \mid r.\mu \neq \text{navigate} \} \\
\weff{\att{XSL}}{\pol{RIP}} &= \{ r \in \att{XSL} \mid r.\mu = \text{navigate} \} \\
\weff{\att{XSL}}{\pol{SIP}} &= \emptyset
\end{align*}
These residuals recover the exact XS-Leak inclusion method taxonomy from the literature~\cite{SoK}, providing a direct mapping between policies and XS-Leak classes. The FIP and NIP cover \emph{frame}-based XS-Leaks (embedded via \code{<iframe>}, \code{<object>}, \code{<embed>}), the NIP covers \emph{window}-based leaks (\code{window.open}) and the RIP covers \emph{direct} leaks (via HTML elements, \code{XMLHttpRequest}, \code{fetch}). The SIP blocks all cross-site requests, therefore providing complete protection against XS-Leaks.

\bfparagraph{Clickjacking.}
Clickjacking attacks require that the target website is loaded on a malicious website through an iframe. 
\begin{align*}
\att{CJ} = \{ &r \in \mathcal{F} \mid r.\sigma = \text{cross-site} \,\, \land \\
&r.\delta \in \demb \land r.\mu = \text{navigate} \}
\end{align*}
\textit{Isolation Policy Effectiveness:}
\begin{align*}
\forall P \in \{\pol{FIP}, \pol{NIP}, \pol{SIP}\}, \,\, &\weff{\att{CJ}}{P} = \emptyset \\
&\weff{\att{CJ}}{\pol{RIP}} = \att{CJ}
\end{align*}
FIP, NIP, and SIP provide complete protection. For all $r \in \att{CJ}$, $r.\sigma = \text{cross-site}$ and $r.\mu = \text{navigate}$, implying $\pol{FIP}(r) = \bot$, $\pol{NIP}(r) = \bot$, and $\pol{SIP}(r) = \bot$. RIP offers no mitigation since it allows cross-site navigational GET requests.

\subsection{Discussion}
\label{sub:policy-effectiveness-discussion}

Among the standard policies, only the SIP fully covers all four attack classes. However, if deployed site-wide, this presents a significant cost to functionality since it disables all cross-site interactions. The NIP offers protection similar to the FIP, since requests that load frames are considered \emph{navigational}. However, the NIP has a narrower scope, given that it blocks only cross-site navigations, whereas the FIP does not consider the site relation. The RIP is the broadest of the non-SIP policies and provides a good balance between security and functionality. However, because it allows cross-site navigations, it is ineffective against attacks such as Clickjacking.

As noted, we scope CSRF to state-changing requests that use unsafe methods, following the HTTP safe-method semantics~\cite{safe_methods}. In practice, it has been reported that around $10\%$ of state-changing operations on popular websites are performed through GET requests~\cite{StateOfSameSite}, violating safe-method semantics. Since state-changing GET requests are spec non-compliant, we omit them from our formal definition of CSRF.
If state-changing GET requests are an application-level choice, then a generic request-context policy like the RIP is not appropriate.
Still, the RIP remains effective for a subset of GET CSRF requests that are non-navigational (e.g., subresource loads, fetch requests). The remaining gap is shared by \code{SameSite=Lax} cookies, the browser default for CSRF protection~\cite{SameSite}, since \code{Lax} cookies are also attached to top-level cross-site navigations.

Taint-style vulnerabilities arise where attacker-controlled data from a given \emph{source} flows into a sensitive \emph{sink}~\cite{pixy}, leading to a range of attacks~\cite{robbery} that includes Cross-Site Scripting (XSS), client-side CSRF~\cite{jaw} and DOM-based open redirects~\cite{open_redirect}. We exclude these attacks from our analysis because (1) they are typically client-side only with no server visibility, and (2) they are delivered to the victim through phishing-style direct navigations (a user clicks a link in an email) rather than cross-site requests, making them indistinguishable from legitimate requests from the server's perspective. However, when malicious content is accessed through cross-site navigation (e.g., an attacker website \emph{opens} a pop-up to a vulnerable cross-site page), FM policies offer some incidental protection.

\section{Systematic Evaluation of Fetch Metadata Against Other Defenses}
\label{chap:fetchComparison}
In this section, we perform the first formal comparison between Fetch Metadata and other security mechanisms to identify gaps, overlaps, and potential security implications caused by differences in enforcement modes (client-side vs. server-side). We provide a proof-of-concept (PoC) of the attacks described in this section in Appenxid~\ref{appendix:poc} and in our artifact~\cite{repo}. We consult documentation resources~\cite{mdn, XSLeaks} to compile a list of security mechanisms that defend against the attack classes covered in Sec.~\ref{chap:isolationPolicies}. We then leverage our formal model to analyze the overlap between FM and each mechanism, identifying if and how they can be covered by FM policies.

First, we define additional formal concepts to help us reason about the enforcement position and timing of security policies during a request's lifecycle.

\bfparagraph{Enforcement Position.} We model a security mechanism $M$ by its set of allowed requests $\{ r \in \mathcal{R}_B \mid P_M(r) = \top \}$ and its  enforcement position $\tau(M) \in \{t_{pre}, t_{serv}, t_{post}\}$. $\tau(M)$ can occur in one of three stages in the request lifecycle: 
\begin{description}[leftmargin=0pt, labelindent=0pt, itemsep=0pt]
    \item \emph{Client-side pre-request} ($t_{pre}$): before the browser sends a request. For example, a request initiated by a \code{<script>} tag is blocked because the host is not allowed by the Content Security Policy (CSP).
    \item \emph{Client-side post-response} ($t_{post}$): after the browser receives the response but before the response content is loaded into the page. For example, a script execution is blocked because the hash of the script is not in the CSP allowlist.
    \item \emph{Server-side request processing} ($t_{serv}$): after the browser sends the request ($t_{pre}$) but before the response is sent to the browser.
\end{description}

\bfparagraph{Enforcement Gap.}
Mechanisms enforced at $t_{post}$ allow for server-side computation before enforcement, affecting response timing. At $t_{post}$, observable timing $\Delta t_{obs} \approx \Delta t_c(s) + \Delta_{\scriptstyle \text{net}}$, where $\Delta t_c(s)$ represents server-side computation dependent on user state $s$ and $\Delta_{\scriptstyle \text{net}}$ is the network round-trip time. Some XS-Leaks rely on measuring timing differences in server-side computation dependent on user data. Historically, this side channel has been difficult to exploit in the wild due to high noise levels (e.g., inconsistent $\Delta_{\scriptstyle \text{net}}$ or $\Delta_{\scriptstyle \text{net}} \gg \Delta t_c(s)$). However, exploiting these discrepancies has recently become easier by leveraging HTTP/2 request and response multiplexing in \textit{timeless} timing attacks \cite{timeless}. 
Mechanisms enforced at $t_{serv}$ (e.g., FM policies) can reject requests before any server-side computation dependent on user state occurs ($\Delta t_c = 0$), resulting in $\Delta t_{obs} \approx \Delta_{\scriptstyle \text{net}}$. The FM specification \cite{west_20_fetch} and other resources~\cite{weichselbaum_2020_protect} warn about timing attacks based on server-side processing and how FM can mitigate this issue by rejecting requests without processing their contents.

\bfparagraph{Mechanism Overlap.}
For each mechanism $M$, we define its \emph{policy context} as the set of request properties it can use to express policies: $V_M \subseteq \{ \sigma, m, \mu, \delta, \text{cookies}, \dots \}$. 

Let $\mathcal{P}_M$ represent the space of all policies expressible using $V_M$. We define two policies $P_1$ and $P_2$ as \emph{equivalent} if they yield identical decisions for every request ($\forall r \in \mathcal{R}, P_1(r) = P_2(r)$). The overlap between mechanisms $M_1$ and $M_2$ is a function of their policy contexts over the request space $\mathcal{R}$:
\begin{description}[leftmargin=0pt, labelindent=0pt, itemsep=0pt]
    \item \emph{Full Overlap}. $V_{M_1} \subseteq V_{M_2}$. It is possible to replicate any policy expressible by $M_1$ using $V_{M_2}$: 
    $$ \forall P \in \mathcal{P}_{M_1}, \exists P' \in \mathcal{P}_{M_2} \text{ s.t. } \forall r \in \mathcal{R}, P'(r) = P(r) $$
    
    \item \emph{Partial Overlap}. $V_{M_1} \cap V_{M_2} \neq \emptyset$ and $V_{M_1} \not\subseteq V_{M_2}$. Mechanism $M_1$ relies on context unavailable to $M_2$. Therefore, $M_1$ can distinguish between requests that appear identical to $M_2$, allowing it to construct policies that cannot be replicated:
    $$\exists P \in \mathcal{P}_{M_1} \text{ s.t. } \forall P' \in \mathcal{P}_{M_2}, \exists r \in \mathcal{R}, P'(r) \neq P(r)$$
    
    \item \emph{No Overlap.} $V_{M_1} \cap V_{M_2} = \emptyset$. Mechanism $M_1$ shares no policy context with $M_2$.
\end{description}

We now compare Fetch Metadata against existing security mechanisms. Tab.~\ref{tab:FMvsProtectionMechanisms} summarizes the overlap and enforcement points. For every attack identified as a result of an enforcement gap, we provide a proof-of-concept attack in Appendix~\ref{appendix:poc}.

\begin{table}[!ht]
\centering\small
\caption{Fetch Metadata vs. other protection mechanisms.}
\begin{tabular*}{\columnwidth}{l@{\extracolsep{\fill}}lcc}
\toprule
& \textbf{Security Mechanism} & \textbf{$\tau$} & \textbf{Overlap} \\ 
\midrule
\multirow{2}{*}{Server} 
& Origin Header   & $t_{serv}$ & \faAdjust \\ 
& Referer Header & $t_{serv}$ & \faAdjust \\ 
\midrule
\multirow{6}{*}{Client} 
& X-Frame-Options & $t_{post}$ & \faAdjust \\
& CSP (\code{frame-ancestors}) & $t_{post}$ & \faAdjust \\ 
& SameSite Cookies & $t_{pre,serv}$ & \faAdjust \\ 
& CORP & $t_{post}$ & \faCircle \\ 
& COOP & $t_{post}$ & \faAdjust \\ 
& X-Content-Type-Options & $t_{post}$ & \faCircle \\ 
\bottomrule
\end{tabular*}
\vspace{2pt}
\caption*{\faCircle~Fetch Metadata can replace the mechanism \faAdjust~Fetch Metadata offers partial coverage of the mechanism}
\label{tab:FMvsProtectionMechanisms}
\end{table}

\subsection{Server-Side Enforced}
\label{subsec:server-mechanisms}

Servers naturally have access to the HTTP method ($m$) of a request. Therefore, we treat $m$ as a baseline request property for all policies enforced at $t_{serv}$. Additionally, we identify $K$ as the broader set of properties revelant to security policies available at server-side. These can be properties of requested resoures, such as their MIME type. In the following, we exclude $m$ and $K$ from the policy context $V_M$ of each server-side enforced mechanism, including FM, so as to not add artificially overlap between mechanisms. They are nonetheless part of the policy context of FM which we define as $V_{FM} \subseteq \{ \sigma, m, \mu, \delta\} \cup K$.

\bfparagraph{Origin Header.} 
The \code{Origin} header is automatically added to requests and contains the origin (scheme, hostname, port) of the request initiator. A web server can then reject requests based on their origin, potentially mitigating cross-origin and cross-site attacks such as those described in Sec.~\ref{sub:protectionAgainstAttacks}. However, browsers do not attach the \code{Origin} header to all requests. In particular, same-origin \code{GET} or \code{HEAD} requests (e.g., performed typing the URL in the omnibox or following a link)~\cite{MDN23_origin} or cross-origin GET requests made in \code{no-cors} mode~\cite{MDN23_origin} do not include the \code{Origin} header. Additionally, the \code{Origin} header can be set to \code{null} for a number of cases such as the use of \code{sandbox} directives~\cite{MDN23_origin}. In all of these cases, the server is unable to distinguish cross-origin from same-origin requests.

We define the policy context of the Origin header as $V_{\text{Origin}} = \{ \text{origin} \}$. Because a server can infer the site relation ($\sigma$) from the specific origin, $V_{\text{Origin}}$ intersects with $V_{\text{FM}}$ ($V_{\scriptstyle \text{FM}} \cap V_{\scriptstyle \text{Origin}} \neq \emptyset$). However, this inferred $\sigma$ is weaker than its FM counterpart, since the \code{Origin} header may not be present. While FM provides a broader policy context ($V_{\scriptstyle \text{FM}} \not\subseteq V_{\scriptstyle \text{Origin}}$), the Origin header provides granular identity that FM abstracts away ($V_{\scriptstyle \text{Origin}} \not\subseteq V_{\scriptstyle \text{FM}}$), implying a \emph{partial overlap} according to our definition.

\bfparagraph{Referer Header.} 
The \code{Referer} header is similar to the \code{Origin} header, but it can additionally contain the path and query string. This extra information allows for even finer-grained policies. However, it can also contain sensitive user data that can be leaked or used for tracking, unlike Fetch Metadata. The content of the \code{Referer} header can be configured through a referrer policy~\cite{referrer_policy} to, for example, send only the origin of a request or even turn off the header entirely. In contrast to the \code{Origin} header, browsers attach the \code{Referer} header to all requests unless explicitly disabled or when the user visits the page directly (i.e., when the navigation does not originate from another page). We define its policy context as $V_{\text{Referer}} = \{ \text{referrer-url} \}$. For the same reasoning applied to the \code{Origin} header ($V_{\scriptstyle \text{FM}} \not\subseteq V_{\scriptstyle \text{Referer}}$) and ($V_{\scriptstyle \text{FM}} \cap V_{\scriptstyle \text{Referer}} \neq \emptyset$), the \code{Referer} header \emph{partially overlaps} with Fetch Metadata.

\subsection{Client-Side Enforced}
\label{subsec:client-mechanisms}

\bfparagraph{X-Frame-Options (XFO).} The \code{X-Frame-Options} response header indicates whether a browser is allowed to render a page in a \code{<iframe>}, \code{<frame>}, \code{<embed>} or \code{<object>} element. This mechanism mitigates attack flows that rely on embedding content from the target website, such as Clickjacking and frame-based XS-Leaks. The header can have two possible values: \code{DENY} -- the page cannot be rendered in a frame; \code{SAMEORIGIN} -- the page can only be rendered in a frame that is same-origin with the requested page. The XFO specification~\cite{xfo} warns that implementations of these criteria are subject to interpretation. We empirically verify that modern browsers (Chromium, Safari and Firefox) consider a frame request if all origins in the frame's ancestor chain (i.e., the parent chain up to the top-level page) are same-origin with the requested page. The code for this experiment is available in the artifact~\cite{repo}.
For XFO, $V_{\scriptstyle \text{XFO}} = \{ \delta \in \demb, \text{anc} \}$, where $\text{\emph{anc}}$ denotes the origin relation of the ancestor chain, and $\tau(\text{XFO}) = t_{post}$.

\emph{Overlap.} $V_{\scriptstyle \text{XFO}} \not\subseteq V_{\scriptstyle \text{FM}}$: the ancestor chain $\text{\emph{anc}}$ cannot be expressed using \code{Sec-Fetch-*} headers and is not available to the server through the other expected request headers. Practically, FM policies alone cannot replicate every behavior achievable with XFO, therefore they \emph{partially overlap}. A block-all FIP fully replicates \code{X-Frame-Options: DENY}, rejecting all requests to embedded contexts. However, \code{SAMEORIGIN} enables a scenario where an embed request is rejected whose immediate parent is same-origin but a cross-origin ancestor exists in the chain. Concretely, consider origins \code{A} and \code{B} and the framing scheme \code{ABB}, where \code{A} is the top-level page, which embeds a page from origin \code{B}, which itself embeds another page from \code{B}. XFO \code{SAMEORIGIN} blocks the innermost frame while allowing the transition frame from \code{A} to \code{B}. A per-request FM policy cannot replicate this behavior. However, a site-wide FIP-like policy would offer the same security guarantees, if the goal is to prevent the innermost frame from ever being loaded, since it would block the transition frame (from \code{A} to \code{B}). 

\emph{Enforcement Gap.} $\tau(\text{XFO}) = t_{post}$, while $\tau(\text{FM}) = t_{serv}$. XFO enforcement occurs after potential stateful computation $\Delta t_c(s) \geq 0$, enabling timing XS-Leaks (PoC in Appendix~\ref{appendix:poc}). $\mathcal{P}_{\scriptstyle \text{FM}}$ can be enforced before the request content is processed ($\Delta t_c = 0$), eliminating the side-channel.

\bfparagraph{Content Security Policy (CSP).} CSP is a broad security standard that allows websites to define directives to control, among others, which resources can be loaded or executed on a given page, helping to prevent attacks such as XSS and Clickjacking. The \code{frame-ancestors} directive specifies a list of valid parents that are allowed to embed a page using \code{<frame>}, \code{<iframe>}, \code{<embed>}, or \code{<object>}, overriding the XFO header if present. For CSP, $\tau(\text{frame}) = t_{post}$ and $V_{\scriptstyle \text{frame}} = \{ \delta \in \demb, \text{origin}, \text{anc}, \sigma \in \{\text{same-origin}\} \}$.

\emph{Overlap.} $V_{\scriptstyle \text{frame}} \cap V_{\scriptstyle \text{FM}} \neq \emptyset$, due to the $\delta$ and $\sigma$ available to both mechanisms. Additionally, $V_{\scriptstyle \text{frame}} \not\subseteq V_{\scriptstyle \text{FM}}$, since CSP \code{frame-ancestors} allows for policies that reason about the ancestor chain \text{\emph{anc}} and allowed origins, which are not available to FM policies. On the other hand, $\delta$ and $\sigma$ are constrained for $M_{\scriptstyle \text{frame}}$, so $V_{\scriptstyle \text{FM}} \not\subseteq V_{\scriptstyle \text{frame}}$. Thus, Fetch Metadata partially overlaps with $M_{\scriptstyle \text{frame}}$.

\emph{Enforcement Gap.} $\tau(\text{frame}) = t_{post}$, therefore enabling a similar timing side-channel attack as XFO (PoC in Appendix~\ref{appendix:poc}).

\bfparagraph{SameSite Cookies.} The SameSite cookie attribute defines three cookie access policies: \code{None}: cookie always sent; \code{Lax}: cookie not sent on cross-site requests except for top-level navigations; \code{Strict}: cookie only sent on same-site requests or direct navigations. Unlike pure client-side or server-side mechanisms, SameSite enforcement has two phases ($\tau(\text{SameSite}) = \{t_{pre}, t_{serv}\}$). At $t_{pre}$, the browser decides whether to attach the cookie based on the SameSite policy. The server ($t_{serv}$) then decides how to process the request based on the cookie's presence and the application logic. 
Because \code{Lax} policies differentiates top-level navigations from subresource fetches, the mechanism relies on mode ($\mu$) and destination ($\delta$). Therefore, $V_{\scriptstyle \text{SameSite}} = \{ \sigma, \mu, \delta, \text{cookie} \}$, where $\sigma \in \{\text{same-site}, \text{none}\}$, $\mu \in \{\text{navigation}\}$ and $\delta \in \{\text{document}\}$.

\emph{Overlap.} $V_{\scriptstyle \text{SameSite}} \cap V_{\scriptstyle \text{FM}} \neq \emptyset$ due to shared (but constrained) $\sigma$, $\mu$ and $\delta$. However, $V_{\scriptstyle \text{SameSite}} \not\subseteq V_{\scriptstyle \text{FM}}$, since it is possible to implement cookie-specific policies (unavailable to FM). Thus, FM \emph{partially overlaps} with the SameSite cookie attribute. Fetch Metadata can mimic policies based on SameSite cookies that only reason about its presence to infer $\sigma$, $\mu$ and $\delta$.

\bfparagraph{Cross-Origin Resource Policy (CORP).} CORP is an opt-in mechanism that restricts access to cross-origin resources imlemented through a response header. It has three possible values: \code{same-origin}, \code{same-site}, and \code{cross-origin}, used to dictate which resources can be accessed by a page: those loaded from the same origin, same site or from any origin, respectively. 
When a \code{no-cors} request is made (e.g., loading an image in an \code{<img>} tag), the browser checks the CORP header to determine if the resource can be accessed by the page, preventing cross-origin information leaks. CORP does not apply to navigational requests. For CORP, $V_{\text{CORP}} = \{ \mu, \sigma \}$ and $\tau(\text{CORP}) = t_{post}$.

\emph{Overlap.} $V_{\scriptstyle \text{CORP}} \subseteq V_{\scriptstyle \text{FM}}$, therefore Fetch Metadata \emph{fully overlaps} with the CORP. It can replace it through a custom policy that rejects \code{no-cors} requests violating \code{same-site}, \code{same-origin}, or \code{cross-origin} boundaries. 

\emph{Enforcement Gap.} $\tau(\text{CORP}) = t_{post}$, therefore enabling timing side-channel attacks (PoC in Appendix~\ref{appendix:poc}).

\bfparagraph{Cross-Origin Opener Policy (COOP).} 
COOP is a security mechanism aimed at mitigating XS-Leaks. It restricts access to top-level documents navigated to using \code{window.open} by separating them in a new browsing context. In practice, it restricts access to window handler fields like \code{length}, \code{location}, and \code{origin}, and sets the \code{window.opener} property in the opened page to \code{null}. We refer to this separation as \emph{isolation}. There are four values for the COOP header: \code{unsafe-none} (default value) turns off COOP; \code{same-origin-allow-popups} isolates documents unless their opener has the same COOP and is same-origin; \code{same-origin} is similar, but the opened page must only contain same-origin documents; \code{noopener-allow-popups} always isolates the document, regardless of the origin of its opener.
For COOP, $V_{M_{\scriptstyle \text{COOP}}} = \{ \sigma \in \{\text{same-origin}\}, \text{opener} \}$ and $\tau(M_{\scriptstyle \text{COOP}}) = t_{post}$.

\emph{Overlap.} $V_{\scriptstyle \text{COOP}} \cap V_{\scriptstyle \text{FM}} \neq \emptyset$ due to restricted $\sigma$. However, $V_{\scriptstyle \text{COOP}} \not\subseteq V_{\scriptstyle \text{FM}}$ (opener relationship unavailable to FM) and $V_{\scriptstyle \text{FM}} \not\subseteq V_{\scriptstyle \text{COOP}}$ ($m, \mu, \delta$ unavailable to COOP). COOP enforces client-side script isolation that server-side FM cannot express. Conversely, FM provides request-level filtering that COOP cannot enforce. While COOP prevents window-handler leaks (e.g., frame counting), it enables cross-window timing attacks that FM can block~\cite{repo}. Therefore, Fetch Metadata \emph{partially overlaps} with COOP.

\emph{Enforcement Gap.} $\tau(\text{COOP}) = t_{post}$, so response timing attacks are enabled (PoC in Appendix~\ref{appendix:poc}).

\bfparagraph{X-Content-Type-Options (XCTO).} XCTO is a response header that signals the browser to respect the MIME type specified in the Content-Type header, preventing MIME type sniffing. XCTO only applies to requests destined to \code{<script>} or \code{<style>} elements. 
XCTO mitigates some XSSI attacks. For example, if an attacker attempts to load \code{application/octet-stream} in a \code{<script>} tag without XCTO, the browser interprets it as JavaScript.
For XCTO, $V_{\text{XCTO}} = \{ \delta \in \{\text{script}, \text{style}\}, \text{MIME} \}$ and $\tau(\text{XCTO}) = t_{post}$.

\emph{Overlap.} $V_{\scriptstyle \text{XCTO}} \subseteq V_{\scriptstyle \text{FM}}$, since Fetch Metadata has a broader $\delta$, and the resource's MIME type is included in $K$ which is part of $V_{\text{FM}}$. Therefore, Fetch Metadata \emph{fully overlaps} with XCTO. Using Fetch Metadata, we can define a policy that checks if the MIME type of the requested resource is compatible with the target element by using the \code{Sec-Fetch-Dest} header and the server's knowledge of the resource. 

\emph{Enforcement Gap.} $\tau(\text{XCTO}) = t_{post}$, so XCTO enables timing attacks via \code{onload}/\code{onerror} handlers similar to the other $t_{post}$ mechanisms (PoC in Appendix~\ref{appendix:poc}).

\section{Testing-Based Policy Inference Methodology}
\label{chap:methodology}
\begin{figure}[t!]
    \centering
    \includegraphics[width=0.9\columnwidth]{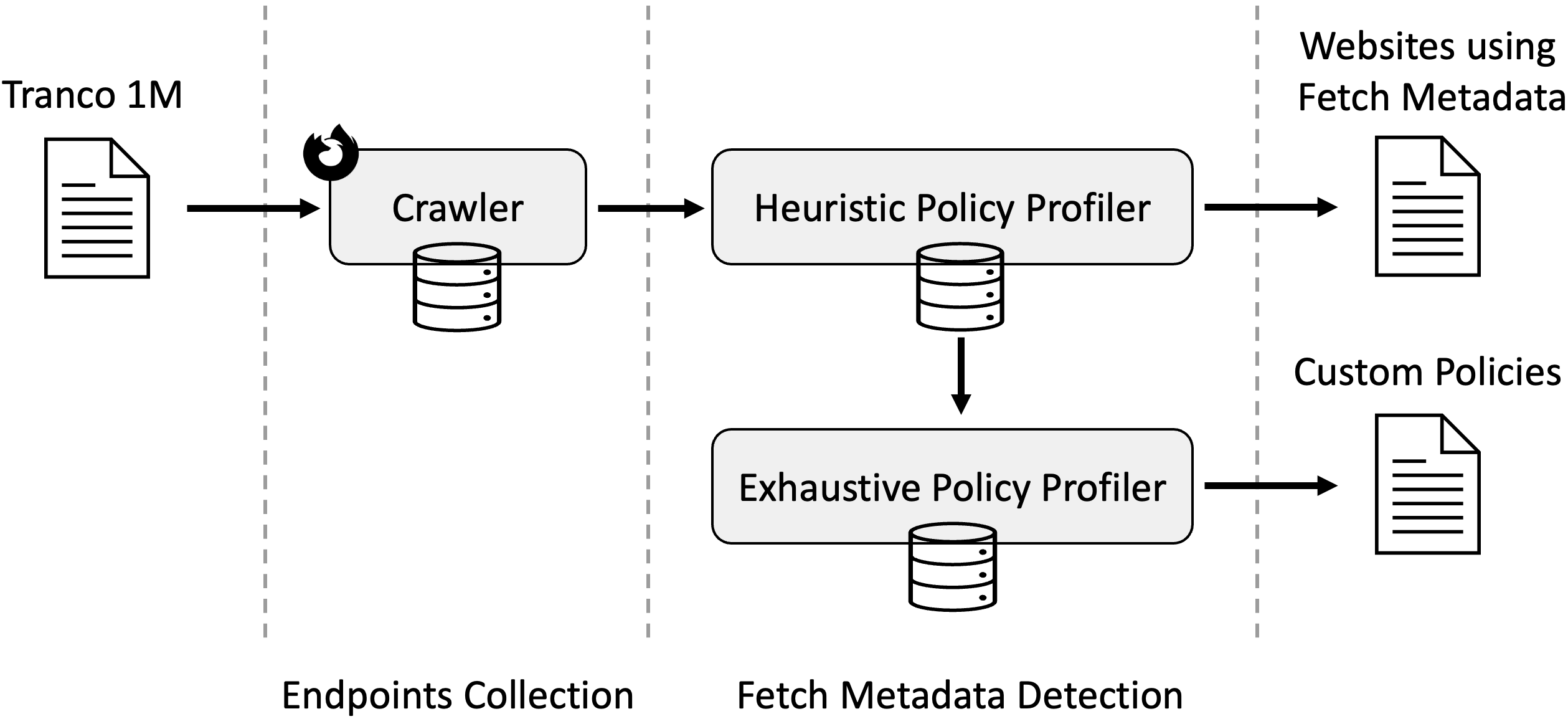}
    \caption{Fetch Metadata detection pipeline.}
    \label{fig:pipeline}
\end{figure}

In this section, we introduce our measurement methodology, illustrated by the pipeline in Figure~\ref{fig:pipeline}. This section describes the three primary components: the crawler, and the heuristic and exhaustive policy profilers, providing a comprehensive overview of each. We also performed a systematic study on the support of Fetch Metadata policies in popular open source server-side frameworks, CMSes and webservers. We refer to the insights gained from this study throughout this section and provide the details in Appendix~\ref{chap:server-side-software}.

\subsection{Endpoint Collection}
\label{subsec:endpoint_collection}
Fetch Metadata policies can be enforced on individual endpoints, so detecting their deployment requires analyzing a representative set of endpoints per website. We develop a crawler using Playwright's Firefox browser in headless mode to visit the landing page of each target website, and extract all endpoints. From these, we focus on those most likely to be protected by FM policies, guided by our expert knowledge, insights from our policy-support study (Appendix~\ref{chap:server-side-software}), and online resources~\cite{west_20_fetch, mdn, Appspot}.

Endpoints referenced in \code{<form>} tags are typically associated with state-changing operations (e.g., login or logout) and are therefore more likely to be protected against attacks like CSRF. We also included requests made during page load, as well as those referenced by anchor tags and frames. While these may not directly trigger state changing operations, they can still be protected by FM policies designed to mitigate threats such as Clickjacking and XS-Leaks. Furthermore, we stored cookies, form parameters, and CSRF-Tokens relevant for each endpoint. 
This information helps us replicate the original requests as close as possible during our measurement, reducing potential inconsistencies that can cause requests to be rejected by the server. 
We identify tokens using the methodology described in~\cite{SecurityTokensReusage}. Note that we only consider same-site endpoints for each crawled page.
Next, we exclude resource names from the URLs collected, keeping only the path components since FM policies are typically applied at the directory level (see Appendix~\ref{chap:server-side-software}). As an exception, we keep the entire URL for specific file extensions that serve HTML content, namely \code{.html}, \code{.htm}, \code{.php}, \code{.asp}, and \code{.aspx}. These resources are more likely to be subject to FM policies depending on the content they deliver and the functionalities they provide. Additionally, for each website, we generate a random URL that corresponds to a non-existent page, e.g., \url{https://example.com/dztwbpugji}, as we observed that websites often implement specific policies for error pages, including \code{404} errors. After, we randomly select up to 50 endpoints per website to use in the next stage of the pipeline.

\subsection{Fetch Metadata Policy Inference}
Besides determining whether the collected endpoints are protected by FM policies, the goal of the study is the identification of the specific policies in place.
We developed two testing tasks to automatically detect and identify FM policies on the collected endpoints: the \LightTester~(\lighttester) is designed to detect if a website enforces Fetch Metadata and identify common policies presented in Sec.~\ref{chap:isolationPolicies} and all their combinations. After this stage, if we cannot determine which policy is in place, we run the \ThoroughTester~(\thoroughtester) that tests all possible combinations of Fetch Metadata headers to identify custom or lesser-known policies. We attach the appropriate cookies, CSRF-Tokens, and form parameters for every request in both profilers. We rename form parameters before sending each request. The goal is to avoid triggering state-changing operations on the server while not affecting FM policy enforcement (see Appendix~\ref{chap:server-side-software}). We acknowledge that renaming parameters may cause validation errors that could be logged or generate notifications. However, we consider this an acceptable risk compared to the alternative of submitting unchanged form parameters and causing state changes.

\subsubsection{Policy Detection Oracle} 

Unlike client-side security mechanisms that are typically enabled through HTTP headers, FM policies are enforced on the server, making them invisible to the browser. To determine whether an FM policy is being enforced on a given endpoint, we require an oracle. This oracle must give us the ability to detect differences in endpoint behavior for specific combinations of \code{Sec-Fetch-*} headers. The natural approach is to use differences in HTTP responses as the oracle. To do so, we first need to establish a baseline that represents the expected behavior of the endpoint when no FM policy is active. However, responses can differ in several ways that do not necessarily imply that a policy is being enforced. Through manual testing, we identify a set of response features that can be used to compare responses while reducing the risk of false positives. In particular, we leverage differences in the HTTP status code and in the response body. If the HTTP status code differs from the baseline, we conclude that a FM policy was violated. If the status code matches the baseline, we then compare the HTML structure of the response body. If there is a significant deviation from the baseline, we conclude that a FM policy is enforced. This approach can still lead to false positives due to dynamic content, but it is less sensitive than string comparison.

\subsubsection{\LightTester~(\lighttester)} 
The \lighttester~sends the following 28 requests for each endpoint: 10 requests to establish the endpoint behavior baseline; 8 requests with different values for the Fetch Metadata headers to trigger policy violations; 10 additional requests to confirm that endpoint behavior continues to match the baseline.

We start by sending 10 requests, 5 POST and 5 GET, with the following headers: \code{Sec-Fetch-Site: same-origin}, \code{Sec-Fetch-Mode: navigate}, and \code{Sec-Fetch-Dest: document}. We chose these Fetch Metadata header values as a baseline because they match the values sent on same-origin top-level navigations, e.g., from \url{https://example.com} to \url{https://example.com/foo} caused by a user clicking on a link. We consider the responses to these requests to be the normal behavior for that endpoint, i.e., how the endpoint should react to legitimate requests. We assume these requests do not violate any FM policy, making them a suitable baseline. We purposely decided not to omit the FM headers from the baseline requests since some websites may refuse them. In that case, we would not be able to detect the presence of FM policies on those endpoints. We assume that the endpoints respond to both GET and POST requests and discard endpoints that do not. This decision streamlines our process and avoids special cases, ensuring correct behavior of our policy detection method.
We opted to make 10 baseline requests to ensure that the endpoint behaves consistently for identical requests. This approach gives us a clear idea of the stability of an endpoint when accessed repeatedly.

Since we know which headers are rejected by the different policies (see Sec.~\ref{chap:isolationPolicies}), we can infer the policy from the responses to specifically crafted combinations of \code{Sec-Fetch-*} headers and HTTP request method. Tab.~\ref{tab:RequestsHeadersVsPolicies} shows the requests used to test the policies in effect for a given endpoint. The column \code{Base Method} indicates the HTTP method (GET or POST) originally associated with the endpoint.
We do not use the \code{Sec-Fetch-User} header since the target policies do not consider it. Additionally, preliminary manual tests on various endpoints revealed that this header does not affect the behavior of the endpoints.

\begin{table*}[!ht]
\small
\caption{Fetch Metadata policy behavior for different \code{Sec-Fetch-*} header combinations.}
\resizebox{\textwidth}{!}{
\begin{tabular}{c|c|c|c|c|c|c|c|c|c|c|c|c|c|c|c}
\toprule
\textbf{Base Method} & \textbf{Method} & \textbf{Sec-Fetch-Site} & \textbf{Sec-Fetch-Mode} & \textbf{Sec-Fetch-Dest} & \textbf{RIP} & \textbf{NIP} &  \textbf{SIP} &\textbf{FIP} & \textbf{rFIP} & \textbf{FIP+RIP} & \textbf{rFIP+RIP} & \textbf{FIP+NIP} & \textbf{rFIP+NIP} & \textbf{FIP+SIP} & \textbf{rFIP+SIP} \\
\midrule
\multirow{8}{*}{GET} & GET  & cross-site & cors & image & \faBolt{} & \faCheck{} &  \faBolt{} &\faCheck{} & \faCheck{} & \faBolt{} & \faBolt{} & \faCheck{} & \faCheck{} & \faBolt{} & \faBolt{} \\
& POST & cross-site & cors & image & \faBolt{} & \faCheck{} &  \faBolt{} &\faCheck{} & \faCheck{} & \faBolt{} & \faBolt{} & \faCheck{} & \faCheck{} & \faBolt{} & \faBolt{} \\
& GET  & cross-site & navigate & image & \faCheck{} & \faBolt{} &  \faBolt{} &\faCheck{} & \faCheck{} & \faCheck{} & \faCheck{} & \faBolt{} & \faBolt{} & \faBolt{} & \faBolt{} \\
& GET  & cross-site & navigate & document & \faCheck{} & \faBolt{} &  \faBolt{} &\faCheck{} & \faCheck{} & \faCheck{} & \faCheck{} & \faBolt{} & \faBolt{} & \faBolt{} & \faBolt{}  \\ 
& GET  & cross-site & navigate & video & \faCheck{} & \faBolt{} &  \faBolt{} &\faCheck{} & \faCheck{} & \faCheck{} & \faCheck{} & \faBolt{} & \faBolt{} & \faBolt{} & \faBolt{}  \\ 
& GET  & same-site & navigate & frame & \faCheck{} & \faCheck{} &  \faCheck{} &\faBolt{} & \faBolt{} & \faBolt{} & \faBolt{} & \faBolt{} & \faBolt{} & \faBolt{} & \faBolt{}  \\ 
& GET  & none & navigate & embed & \faCheck{} & \faCheck{} &  \faCheck{} &\faBolt{} & \faBolt{} & \faBolt{} & \faBolt{} & \faBolt{} & \faBolt{} & \faBolt{} & \faBolt{}  \\
& GET  & same-origin & navigate & frame & \faCheck{} & \faCheck{} &  \faCheck{} &\faBolt{} & \faCheck{} & \faBolt{} & \faCheck{} & \faBolt{} & \faCheck{} & \faBolt{} & \faCheck{}  \\
\midrule
\multirow{8}{*}{POST} & POST  & cross-site & cors & image & \faBolt{} & \faCheck{} &  \faBolt{} &\faCheck{} & \faCheck{} & \faBolt{} & \faBolt{} & \faCheck{} & \faCheck{} & \faBolt{} & \faBolt{}  \\
& POST & cross-site & cors & image & \faBolt{} & \faCheck{} &  \faBolt{} &\faCheck{} & \faCheck{} & \faBolt{} & \faBolt{} & \faCheck{} & \faCheck{} & \faBolt{} & \faBolt{}  \\
& GET  & cross-site & navigate & image & \faCheck{} & \faBolt{} &  \faBolt{}&\faCheck{} & \faCheck{} & \faCheck{} & \faCheck{} & \faBolt{} & \faBolt{} & \faBolt{} & \faBolt{}  \\
& POST  & cross-site & navigate & document & \faBolt{} & \faBolt{} &  \faBolt{} &\faCheck{} & \faCheck{} & \faBolt{} & \faBolt{} & \faBolt{} & \faBolt{} & \faBolt{} & \faBolt{}  \\ 
& POST  & cross-site & navigate & video & \faBolt{} & \faBolt{} &  \faBolt{} &\faCheck{} & \faCheck{} & \faBolt{} & \faBolt{} & \faBolt{} & \faBolt{} & \faBolt{} & \faBolt{}  \\ 
& POST  & same-site & navigate & frame & \faCheck{} & \faCheck{} &  \faCheck{} &\faBolt{} & \faBolt{} & \faBolt{} & \faBolt{} & \faBolt{} & \faBolt{} & \faBolt{} & \faBolt{}  \\ 
& POST  & none & navigate & embed & \faCheck{} & \faCheck{} &  \faCheck{} &\faBolt{} & \faBolt{} & \faBolt{} & \faBolt{} & \faBolt{} & \faBolt{} & \faBolt{} & \faBolt{}   \\
& POST  & same-origin & navigate & frame & \faCheck{} & \faCheck{} &  \faCheck{} &\faBolt{} & \faCheck{} & \faBolt{} & \faCheck{} & \faBolt{} & \faCheck{} & \faBolt{} & \faCheck{}   \\
\midrule
\end{tabular}}
\caption*{\faBolt{} denotes a policy violation under the given method and header combination \faCheck{} denotes no violation}
\vspace{-12pt}
\label{tab:RequestsHeadersVsPolicies}
\end{table*}

Next, we repeat the baseline requests to verify that the endpoint's behavior did not change as a result of the testing. This check helps prune false positives, ruling out endpoints that exhibit different behavior for identical requests. 

We adopt a two-stage testing approach. We start with an initial scan of all endpoints using the \lighttester. For endpoints where Fetch Metadata is detected, we repeat the scan ten additional times. This iterative process allows us to detect transient anomalies and identify endpoints that show consistent policy behavior, significantly improving the reliability of our detection. We reorder the endpoints to prevent endpoints from the same website from being tested concurrently, reducing the computational load on the servers.

\subsubsection{\ThoroughTester~(\thoroughtester)}
When the \lighttester~detects the presence of FM policies but cannot determine the policy in use, we run the \thoroughtester. The \thoroughtester~selects at most two endpoints per website and makes GET and POST requests covering all combinations of Fetch Metadata header values. We exclude the \code{Sec-Fetch-User} header as it is rarely considered in policies, reducing the number of requests needed. To keep the testing light on the server, we reorder endpoints and introduce a $>$2 second delay between requests.

\subsection{Threats to Validity}

The ability to detect the presence of FM policies and infer their behavior is at the core of our pipeline. Therefore, we must ensure our approach is accurate and reliable. A key part of our methodology is the response difference oracle. Detecting changes in endpoint behavior when two responses share the same status code may yield false positives. Endpoints may serve dynamic content that changes with each request, even when FM policies are not violated. As a result, our profilers may detect changes in endpoint behavior, even when the behavior remains consistent with the endpoint's standard operation.
To reduce the impact of inconsistent results, we ran the tests 10 times for all the endpoints where Fetch Metadata was detected in the initial scan and we consider only those endpoints where a FM policy was detected in at least 9 out of the 10 runs. 
For subsequent \thoroughtester~run, we restrict our analysis to endpoints with stable detections across all \lighttester~runs (10/10), as the higher request volume may amplify transient instability observed under \lighttester~measurement.
Furthermore, we did not interpret the status code \code{429 Too Many Requests} as an indication of behavior change caused by Fetch Metadata headers, as this response typically indicates rate limiting rather than a policy violation.

To ensure our oracle is accurate, we conducted a series of experiments on synthetic testbeds and real-world websites. All code is available in our artifact~\cite{repo}.

\bfparagraph{Custom Web Application.} On a minimal Express application with 12 endpoints (GET and POST), each protected by an FM policy from Sec.~\ref{chap:isolationPolicies} or a combination thereof, the \lighttester~identified all policies with 100\% precision and recall.

\bfparagraph{Software Supporting Isolation Policies.} We developed a set of minimal applications using frameworks identified in Appendix~\ref{chap:server-side-software}
(Apache Struts 2, Apache Wicket, Joomla, and Node.js), which offer some out-of-the-box support for FM policies. These policies correspond to the ones described in Sec.~\ref{chap:isolationPolicies} or slight variations. On these applications, the~\lighttester~ detected the presence of a policy in all cases, but could not always determine the exact policy variant. The~\thoroughtester~, however, achieved 100\% precision and recall.

\bfparagraph{Manual Validation.} To establish a ground truth on real websites, we manually altered the values of \code{Sec-Fetch-*} headers using Burp Suite Repeater on a curated list of websites known or strongly suspected to enforce Fetch Metadata (e.g.,~\cite{Appspot} and sites operated by Google). Across this dataset, both profilers achieved 100\% precision and recall in detecting whether FM policies were deployed, with the \thoroughtester~handling both standard and custom policies.

\bfparagraph{Post-Login Analysis.}
Our pipeline is designed to perform the measurement from an unauthenticated position. At first glance, it is sensible to consider that this approach could be less effective. In particular, we identify two main concerns: \cir{1} Endpoints that require authentication are more likely to implement Fetch Metadata isolation policies; \cir{2} It is possible to collect more unique and security-relevant endpoints from an authenticated position. To test these hypotheses, we design a post-login measurement pipeline that performs the same endpoint collection and FM policy detection as our main pipeline but from an authenticated position. We took inspiration from Rautenstrauch et al.~\cite{jannisAuth} that presented a semi-automated login and registration approach featuring a human-in-the-loop for account registration and login. Additionally, we leverage Single Sign-On (SSO) to simplify the registration and login process on the target websites. We use SSO-Monitor Archive~\cite{ssomonitor} to identify the most popular identity provider (IdP) and collect the top 100 websites supporting this IdP. First, we create an account on this IdP (Google). Then, for each website, we navigate to a candidate login page provided by SSO-Monitor and manually register and login using SSO and password auto-fill. After a successful login, we yield control of the browser to the script that collects endpoints as described in Sec. \ref{subsec:endpoint_collection} and runs the \lighttester{} and the \thoroughtester{} if necessary. 
We completed a best-effort measurement which was successful for 61 out of the 100 selected websites. The most common reason for failure was browser automation detection which manifested in different ways across websites. Some websites presented CAPTCHAs during the login process that we could not bypass. For other websites, the login or registration process simply failed without a clear reason. We identified no FM policies deployed on the 61 websites analyzed, providing evidence against \cir{1}. Additionally, the number of unique endpoints collected was lower or equal to the unauthenticated measurement for 79\% of websites, which is evidence against \cir{2}. Considering also that FM policies are recommended to be deployed before any data processing occurs on the server~\cite{weichselbaum_2020_protect}, e.g., session validation, we consider our choice of performing an unauthenticated measurement a good tradeoff between coverage and complexity.

\section{Fetch Metadata Adoption in the Wild}
\label{chap:evaluation}
In this section, we present the results of our large-scale measurement on the adoption of Fetch Metadata among Tranco's~\cite{tranco, LePochat2019} top 1M websites, performed between June 18 and August 16, 2025. 
Tab.\ref{tab:pro_vs_unp} summarizes the results of our scan. Our measurement shows that only 0.23\% of the top 1M websites deploy Fetch Metadata policies to secure at least one of their endpoints. The adoption is higher for more popular websites, reaching 2.8\% for the top 1k sites. Based on \code{WHOIS} data, we attribute 124 and 97 of the identified protected sites to Google and Meta, respectively. Together, they account for 9.6\% of protected sites of the top 1M, and they dominate the top 1k at 92.9\% of protected websites.

\begin{table}[ht]
\centering
\resizebox{\columnwidth}{!}{%
\begin{tabular}{|c|c|c|c|c|}
\hline
\textbf{Top 1k} & \textbf{Top 1k--10k} & \textbf{Top 10k--100k} & \textbf{Top 100k--1M} & \textbf{Overall} \\
\hline
28 [2.8\%] & 40 [0.44\%] & 183 [0.2\%] & 2044 [0.23\%] & 2295 [0.23\%] \\
\hline
\end{tabular}%
}
\caption{Number [and percentage] of protected websites.}
\label{tab:pro_vs_unp}
\end{table}

We identified 14,141 protected endpoints, 12,626 (89.29\%) of which are GET endpoints and 1,515 (10.71\%) POST endpoints. The median reveals that 2.17\% of the tested endpoints on protected sites are secured with FM policies, suggesting that these policies are most often applied selectively. 
Among the protected endpoints, 94\% return a different status code on policy violations, 5.5\% return modified page content and 0.5\% return both: different status codes and page content, depending on the specific header values.

\bfparagraph{Performance.} The average time taken by the \lighttester~per endpoint is 7.4 seconds (median: 4.25s). In contrast, the \thoroughtester~took on average 46 minutes per endpoint (median: 45.1min).

\subsection{Inferred Fetched Metadata Policies}

Tab.~\ref{tab:PoliciesStats} lists the policies identified during the experiments and the number of websites that enforce them. Note that some websites implement more than one policy: 94 websites deploy two policies and 48 websites deploy three or more. Policies that the \lighttester~could not identify are labeled \code{custom}. Following custom policies, the RIP is the most prevalent policy, followed by a combination of the FIP and the RIP. Note that we did not detect any website protected by the NIP, perhaps due to the similar protection profile offered by the FIP, which more directly targets attacks like Clickjacking (see Tab.~\ref{tab:PoliciesVsAttacks}).

\begin{table}[ht]
\small
\centering
\rowcolors{2}{gray!20}{white}
\begin{tabular}{ >{\raggedright\arraybackslash}p{0.65\columnwidth} | r }
\toprule
\rowcolor{white}
\textbf{Policies} & \textbf{Nr. of websites} \\ 
\midrule
\textbf{custom} & 1208 \\ 
\textbf{RIP} [resource] & (52) \quad 877 \\ 
\textbf{FIP} [framing] + \textbf{RIP} & (51) \quad 142 \\ 
\textbf{rFIP + RIP} & (15) \quad 136 \\ 
\textbf{rFIP} & 76 \\ 
\textbf{FIP} & 46 \\ 
\textbf{SIP} [strict] & 22 \\
\textbf{FIP + SIP}  & 8 \\ 
\textbf{rFIP + SIP} & 7 \\ 
\bottomrule
\end{tabular}
\caption{Websites deploying Fetch Metadata policies. The number between () represents a subset of endpoints enforcing the policy only for GET requests.}
\vspace{-12pt}
\label{tab:PoliciesStats}
\end{table}

\bfparagraph{Custom Fetch Metadata Policies}
\label{subsec:resultsCustomPolicies} 
For websites where the \lighttester~could not infer enforced policies (marked as \code{custom} in Tab.~\ref{tab:PoliciesStats}), we performed additional testing using the \thoroughtester. Across all custom policies, we found slight variants of the policies discussed in Sec.~\ref{chap:isolationPolicies}, but also atypical configurations. 
Policy CP1 (Listing~\ref{alg:document-frame-iframe}) is the most common policy, deployed in 244 websites. It allows all POST requests and GET requests targeting \code{document}, \code{frame}, or \code{iframe} contexts, mitigating XSSI. We observed this policy on similarly structured endpoints (e.g., \code{chtm/s/metric}) across sites, which all appear to be hosted on the Chatium cloud platform~\cite{a2026_chatium}.
Policy CP2 (Listing~\ref{alg:cross-site_none}), deployed on 196 websites, blocks all cross-site POST requests and all user-initiated navigations (e.g., clicking on bookmark), thereby mitigating CSRF attacks. Most deployments of this policy occur on forums based on vBulletin~\cite{a2020_vbulletin}.
Policy CP3 (Listing~\ref{alg:cross-site_object_embed}), deployed on 63 websites, blocks cross-site POST requests and cross-site \code{object} and \code{embed} loads, mitigating certain cases of CSRF, Clickjacking, and other cross-site navigation-based attacks.
Policy CP4 (Listing~\ref{alg:image}), deployed on 41 websites, rejects requests initiated via \code{<img>} elements. This policy would break endpoints that serve images, so it is likely meant to mitigate image-based XS-Leaks such as timing or error-based state inference.

\subsection{Fetch Metadata and Security Header Deployment}
FM-deploying sites exhibit a stronger security posture, deploying most security headers at higher rates than the overall top-1M:CSP \code{frame-ancestors}: 53.2\% vs 15.1\%; X-Frame-Options: 88.6\% vs 51.3\%; X-Content-Type-Options: 79.8\% vs 47.7\% (see Tab.~\ref{tab:FM_vs_headers_deployment} in Appendix~\ref{app:headerDeployment}).
 
To understand how websites use security headers alongside FM policies, we classify the FM-protected endpoints into one of four states, related to their deployment of security headers. For each header: \cir{1}~the header is deployed alongside an overlapping FM policy (defense-in-depth); \cir{2}~the header is not present, but an overlapping FM policy is deployed; \cir{3}~the header is deployed alongside a non-overlapping FM policy; \cir{4}~the header is not deployed and there is a non-overlapping FM policy. We say a policy $P$ \emph{overlaps} with header $H$ when they mitigate the same attack class. We associate each header with the attack class (from~\ref{sub:protectionAgainstAttacks}) it mitigates: CSP \code{frame-ancestors} and XFO target clickjacking, frame-based XS-Leaks, CORP targets XSSI and direct XS-Leaks, and COOP targets window-based XS-Leaks. We omit CSRF as its main defenses (SameSite cookies and CSRF tokens) are not headers, and assessing SameSite cookies requires authentication which we deem out-of-scope. Overlapping FM policies are those that mitigate those attack classes. For example, NIP and SIP mitigate window XS-Leaks as discussed in Sec.~\ref{sub:protectionAgainstAttacks} therefore, they overlap with COOP.

\begin{table}[ht]
\footnotesize
\centering
\setlength{\tabcolsep}{4.5pt}
\rowcolors{2}{gray!20}{white}
\begin{tabular}{ >{\raggedright\arraybackslash}p{0.35\columnwidth} | r | r | r }
\toprule
\rowcolor{white}
\textbf{Header} & \textbf{\cir{1}~Both} & \textbf{\cir{2}~FM Only} & \textbf{\cir{3}~Non-Overlap} \\ 
\midrule
\textbf{\code{frame-ancestors}}    & 0.47\% & 12.31\% & 29.22\% \\ 
\textbf{CORP}   & 25.63\% & 16.26\% & 0.85\% \\ 
\textbf{COOP}   & 0.05\% & 2.98\% & 15.21\% \\ 
\textbf{XFO}    & 8.61\% & 4.18\% & 65.57\% \\ 
\bottomrule
\end{tabular}
\caption{Deployment of security headers on FM-protected endpoints in relation to the FM policies deployed.}
\label{tab:FM_vs_headers_distribution}
\end{table}

Tab.~\ref{tab:FM_vs_headers_distribution} reveals two distinct deployment strategies. For resource isolation, developers actively pair CORP with overlapping FM rules, yielding the highest rate of layered defense (25.63\%) and they rarely pair it with disjoint policies (0.85\%). Conversely, framing and navigation controls (\code{frame-ancestors}, COOP, and XFO) rarely overlap with FM ($\le$0.5\% for \code{frame-ancestors} and COOP; 8.61\% for XFO), appearing instead alongside non-overlapping rules (15.21\%--65.57\%). This hints that practitioners primarily use FM to secure subresources and APIs, while relying on header-based defenses for top-level navigation and framing.

\section{Related Work}
\label{chap:relatedWork}
To the best of our knowledge, this paper presents the first systematic analysis of both the effectiveness of Fetch Metadata policies against web attacks and their adoption in the wild. 
In this section, we discuss previous research work conducted in the context of other web security mechanisms.

Several studies focused their attention on the deployment of CSP. In 2014, Weissbacher et al.~\cite{CSP_study} analyzed the adoption of CSP on the top 1M websites of the Alexa ranking over a period of 16 months, and observed that only 815 sites were deploying CSP in enforcement mode. A subsequent study from 2016 by Calzavara et al.~\cite{calzavara_2016_content} similarly reported sparse use of CSP and frequent misconfigurations among the Alexa Top 1M websites. Roth et al.~\cite{roth_2020_complex} used the CSP policies collected by the Internet Archive between 2012 and 2018 to perform a longitudinal study of CSP adoption on 10k highly ranked websites. Although the authors observed a tenfold increase in CSP adoption between 2014 and 2016, they also noted that many sites never enforced the policies, as 56\% of them never switched from report-only mode.

Other large-scale measurements studied the deployment of the SameSite cookie attribute. In~\cite{SameSite}, the researchers measured the attribute usage on the Alexa top 5k websites between January and March 2020, when Chrome started to enforce the \emph{lax by default} policy. In this period, the number of sites not specifying the SameSite attribute dropped from 98\% to 90\%, whereby around 4\% of the websites fully adopted the SameSite attribute, while the other 4\% set the attribute to None, thereby disabling every protection. Khodayari and Pellegrino~\cite{StateOfSameSite} extended~\cite{SameSite} by performing a longitudinal study of 500k sites between June 2019 and March 2021. The study shows a significant deceleration in the adoption rate of SameSite after the completion of the rollout of the \emph{lax by default} policy and that, at the end of the measurement period, only 18.94\% of sites explicitly set a value of the SameSite attribute.

Chen et al.~\cite{corsAnalysis} analyzed in 2018 the usage of CORS headers on the Alexa top 50k websites. The authors observed that CORS was deployed on 22k websites and 481k of their subdomains, and discovered insecure CORS configurations on around 132k subdomains (2.9k base domains).
In 2020, Squarcina et al.\cite{same_site_squarcina} performed a large-scale measurement on the top 50k domains from the Tranco list to quantify the additional threats posed by related-domain attackers to websites with respect to standard web attackers. The paper, which focused on various security-related web mechanisms including cookies, CSP, CORS, postMessage, and domain relaxation, reports that 887 sites were affected by vulnerabilities exploitable only by related-domain attackers.

\section{Discussion and Conclusion}
\label{chap:conclusion}
Our work sheds light on the potential of Fetch Metadata to enhance the security of web applications, while also identifying low adoption rates and barriers that hinder widespread deployment. Fetch Metadata policies provide robust protection against cross-site attacks such as CSRF, XS-Leaks, XSSI and Clickjacking and can strengthen or even replace some popular client-side defenses. Still, our research reveals a low adoption rate of 0.23\% on the Tranco top 1M websites. Two factors help explain this gap. First, the protection offered by Fetch Metadata policies often overlaps with existing and widely deployed client-side mechanisms. These mechanisms are plug-and-play and enforced by the browser, keeping the enforcement responsibility away from the server. Second, our analysis in Sec.~\ref{chap:server-side-software} shows that Fetch Metadata policies lack out-of-the-box support in most popular languages and frameworks. This increases the deployment effort, adding further friction for developers to deploy Fetch Metadata policies. As a first step to confirm our hypotheses and gather new insights from practitioners, we ran a preliminary mixed study~\cite{methodsbook} in the form of an online survey on Prolific~\cite{prolific} about Fetch Metadata policies. However, we found that an online survey is not the right tool to explore this question. The quantitative data obtained is too shallow to extract strong conclusions, and the majority of the open-ended answers yielded low-effort responses or signs of LLM usage. Understanding the concrete reasons behind the low adoption of Fetch Metadata policies calls for a qualitative developer interview-based study, which we leave as future work. 

Fetch Metadata is distinct in that it operates at the server side. This is a tradeoff rather than a pure advantage. Server-side enforcement is particularly well-suited to defend against attacks that rely on successful cross-site or cross-origin requests (CSRF, XS-Leaks, XSSI), or when the defender's goal is to isolate specific resources from cross-origin or cross-site accesses.
Outside of these use cases, the benefits lie mostly in defense-in-depth, adding an extra layer of security when client-side enforcement may be unreliable or too coarse.
Our work formalizes all these considerations and provides a comprehensive reference for developers and researchers, clarifying the security guarantees of different Fetch Metadata isolation policies, the interplay with other security mechanisms, and the gaps left by the difference in enforcement position. This, alongside our server-side policy inference method, provides a foundation for future research not just on Fetch Metadata but on server-side security policies more broadly.

\appendix

\section*{Ethical Considerations}
\label{appendix:ethics}
We adhere to state-of-the-art standards and best practices accepted by the research community \cite{cassel2022omnicrawl, ReproducibilityAndReplicability, webMeasurementsDifferentSetups, SecurityTokensReusage, RedLines, EthicalFrameworks}. We follow the ethical guidelines of leading computer security conferences \cite{UsenixEthics, NDSSEthics}, upholding the principles described in the Menlo Report \cite{Menlo}.

Our study does not actively probe or exploit any server-side vulnerabilities, minimizing the risk of harm to data confidentiality, integrity, and availability. Our large scale measurement approach does not involve authentication flows and authenticated requests, reducing server resource consumption and sensitive data manipulation. We identify ourselves as a part of a research study through custom headers, and generated requests are specifically designed to avoid triggering state-changing actions. Specifically, we rename form parameters before sending requests to avoid triggering state-changing operations while preserving Fetch Metadata policy enforcement; this may cause server-side validation errors, which we consider an acceptable risk as discussed in Sec.~\ref{chap:methodology}. Furthermore, our scanner introduces significant time gaps between requests to avoid affecting service availability. The semi-automated post-login analysis was performed on 100 websites and supervised by one of the authors to ensure that no harmful actions were executed. Finally, another ethical consideration is the potential impact on advertisers if websites serve ads to our crawler. While our experiment may consume some ad budget, we access every website's homepage only once, making this cost negligible.

As discussed in Sec.~\ref{chap:conclusion}, we conducted a preliminary quantitative study with 100 developers through an online survey. We designed this survey according to the guidelines provided by our institutional ethics review board. Participants were informed of the survey's structure, purpose, expected duration and monetary compensation before agreeing to participate. At the start of the survey, we reiterated this information and asked participants for consent before proceeding. We informed participants of their rights under GDPR, including the right to withdraw at any time without penalty, and our GDPR compliant data collection and handling policy. The survey was conducted using a self-hosted platform ensuring full control over the data collected. All data was stored pseudonymously, tied only to each participant's ID in the recruiting platform (Prolific) which is necessary for approving submissions, issuing the payments and collecting non-identifiable demographic information. The median completion time was 59 minutes and 43 seconds, which aligned with our initial duration estimate of 60 minutes advertised to participants. Each participant was paid a flat fee of £12, calculated from the expected duration. The selected hourly rate of £12 was suggested by Prolific as a \emph{great} compensation rate.

Due to the limitations of this preliminary study, we decided to not include the results in the current paper and leave a more in-depth qualitative study with developers as future work.

\bibliographystyle{plainurl}
\bibliography{base}

\section{Assessment of Fetch Metadata Policy Support in Open Source Software}
\label{chap:server-side-software}
The Fetch Metadata policies described in Sec. \ref{chap:isolationPolicies} are straightforward to implement (see Appendix~\ref{appendix:isolationPolicies}). However, their widespread adoption likely depends on the availability of supporting tools and framework integration. Prior research shows that developers prefer using frameworks and libraries to implement security features, as this reduces complexity and helps prevent vulnerable implementations~\cite{hermann_2025_security_features}.

In the following, we study the availability of frameworks and tools that support Fetch Metadata policies. This step is fundamental as it informs the development of our automated testing methodology (see Sec. \ref{chap:methodology}). 

\subsection{Methodology}
\label{appendix:sever-side_methodology}
Our primary goal was to identify server-side software that supports the implementation of Fetch Metadata policies, including built-in features, modules, libraries, tools, and plugins. We used Google and DuckDuckGo to search for keywords such as \textit{Fetch Metadata} and \textit{Resource Isolation Policy}. On GitHub, we looked for repositories that mentioned \textit{Fetch Metadata} in their project descriptions, as well as those containing occurrences of \textit{Sec-Fetch} and \textit{sec-fetch} in the source code, allowing us to identify projects that use these headers. We also reviewed academic and technical literature for relevant discussions. See Table \ref{tab:search} for a detailed list of search terms used. We conducted the most recent search in June 2025. 

\label{appendix:search}
\begin{table}[t]
\small
\centering
\begin{tabular}{>{\centering\arraybackslash}p{\dimexpr\columnwidth-2\tabcolsep}}
\toprule
\textbf{Search terms} \\
\midrule
 \code{Fetch Metadata} \\
 \code{fetch-metadata} \\
 \code{Sec-Fetch} \\
 \code{sec-fetch} \\
 \code{Resource Isolation Policy} \\
 \code{Navigation Isolation Policy} \\
 \code{Frame Isolation Policy} \\
 \code{Strict Isolation Policy} \\
 \code{fetch metadata middleware} \\
 \code{fetch metadata apache} \\
 \code{fetch metadata nginx} \\
 \code{fetch metadata nodeJS} \\
 \code{fetch metadata npm} \\
 \code{fetch metadata laravel} \\
 \code{fetch metadata python} \\
 \code{fetch metadata flask} \\
 \code{fetch metadata django} \\
 \code{fetch metadata WordPress} \\
 \code{fetch metadata Joomla} \\
 \code{fetch metadata Spring Boot} \\
 \bottomrule
\end{tabular}
\caption{Search terms.}
\label{tab:search}
\end{table}

We focused on identifying solutions for the leading open-source web servers, Nginx and Apache, as well as libraries for popular server-side frameworks, including Spring Boot, Laravel, Django, Flask, Node.js, and Struts 2. Additionally, we investigated whether major content management systems (CMSes), such as WordPress and Joomla, offer plugins for Fetch Metadata policies.

\subsection{Results}
\label{appendix:server_side_results}
We identified several frameworks, plugins, and packages that support the implementation of Fetch Metadata policies. These tools typically implement the RIP and allow developers to exempt specific endpoints. We found no corresponding packages or built-in features for Flask, Laravel, Spring Boot, Django, Nginx, and Apache. Although the lack of search results does not confirm the absence of solutions (custom or undocumented implementations may exist) it nevertheless indicates limited support of Fetch Metadata.

\bfparagraph{Apache Struts 2} is an open-source Java web framework. The \textit{Fetch Metadata Interceptor}~\cite{fetchMetadataInterceptor}, developed by Apache, enforces an RIP-like policy on all endpoints by default. It can be activated via the Struts configuration file, with explicit exemptions for endpoints requiring cross-site access. 

\bfparagraph{Apache Wicket} is an open-source, component-based Java web framework. It provides the \textit{FetchMetadataResourceIsolationPolicy} class, which enforces the RIP~\cite{fetchmetadataresourceisolationpolicy} and can be extended to define custom policies.

\bfparagraph{WordPress} is a widely used open-source CMS offering extensive plugin support. The third-party \textit{Fetch Metadata} plugin~\cite{carraro_2020_enricocarrarowpfetchmetadata}, available on GitHub, enables activation of the RIP and NIP. It also supports a report-only mode, allowing developers to monitor policy effects without enforcing isolation.

\bfparagraph{Joomla} is an open-source CMS. Although it lacks native Fetch Metadata support, the third-party \textit{FetchMetadata} extension~\cite{zulauf_2020_fetchmetadata} enables a relaxed version of the RIP across all endpoints, with options to exempt specific ones. It is available on GitHub or the Joomla extension directory.

\bfparagraph{Node.js} is an open-source JavaScript runtime environment for server-side applications. Although it does not natively enforce Fetch Metadata policies, the npm package \textit{fetch-metadata}~\cite{erasmus_2022_fetchmetadata} provides middleware to facilitate their implementation. By default, it enforces a RIP and supports further customization, such as restricting specific \code{Sec-Fetch-Dest} values.

\subsection{Deployment Strategies} 
\label{appendix:server-side_deployment}
We extract several key insights from this manual analysis step that help us design an effective and efficient automated testing methodology. First, Fetch Metadata policies are typically enforced at the directory level. This means that different resources such as \code{.js} files requested on the same endpoint are often subject to the same policy. 

Fetch Metadata policies should be enforced before the application logic processes the request's data, preventing side-channel attacks~\cite{weichselbaum_2020_protect}. This recommendation is reflected in the tool support for Fetch Metadata policies. For example, the Node.js \textit{fetch-metadata}~\cite{erasmus_2022_fetchmetadata} package provides Fetch Metadata policy implementations as a middleware, meaning they will run before the request reaches application code. This observation about Fetch Metadata policy enforcement significantly affects some design choices in our methodology regarding authenticated requests (see Sec.~\ref{chap:methodology}).

Finally, by identifying the policies supported by tools and frameworks, we validate our choice of focusing on the isolation policies described in Sec.~\ref{chap:isolationPolicies}, as we are likely covering most Fetch Metadata policies deployed in the wild.

\section{Security Header Deployment vs. Fetch Metadata in the Wild}
\label{app:headerDeployment}
Websites deploying Fetch Metadata generally exhibit a stronger security posture, as they more frequently enable additional security headers compared to all websites in our dataset (Tab~\ref{tab:FM_vs_headers_deployment}). XFO is an exception, being less frequently deployed among Fetch Metadata-protected sites, which coincides with higher adoption of CSP \code{frame-ancestors} on those websites.

\begin{table}[ht]
\small
\centering
\rowcolors{2}{gray!20}{white}
\begin{tabular}{ >{\raggedright\arraybackslash}p{0.6\columnwidth} | r | r }
\toprule
\rowcolor{white}
\textbf{Security Header} & \textbf{FM Sites} & \textbf{Overall} \\
\midrule
\textbf{CSP (\code{frame-ancestors})} & 53.2\% & 15.1\% \\
\textbf{Cross-Origin-Resource-Policy} & 54.9\% & 10.1\% \\
\textbf{Cross-Origin-Opener-Policy}   & 56.1\% & 11.1\% \\
\textbf{X-Frame-Options}              & 88.6\% & 51.3\% \\
\textbf{X-Content-Type-Options}       & 79.8\% & 47.7\% \\
\bottomrule
\end{tabular}
\caption{Deployment of security headers on websites using Fetch Metadata (FM) compared to the overall set of websites.}
\label{tab:FM_vs_headers_deployment}
\end{table}

\section{Fetch Metadata Isolation Policies}
\label{appendix:isolationPolicies}
This appendix provides pseudocode implementations of the Fetch Metadata isolation policies discussed in Sec.~\ref{chap:isolationPolicies}. Each listing describes the decision logic applied by the corresponding policy.

\begin{center}
\begin{minipage}{0.95\linewidth}
\begin{lstlisting}[caption={Navigation Isolation Policy (NIP).}, label={lst:pseudocode-nip}]
Input: req = HTTP request
if req.headers.secFetchSite = "cross-site" and
   req.headers.secFetchMode = "navigate" then
    reject
allow
\end{lstlisting}
\label{alg:NIP}
\end{minipage}
\end{center}

\begin{center}
\begin{minipage}{0.95\linewidth}
\begin{lstlisting}[caption={Resource Isolation Policy (RIP).}, label={lst:pseudocode-rip}]
Input: req := HTTP request
if req.headers.secFetchSite != "cross-site" then
    allow
if req.headers.secFetchMode = "navigate" and
   req.method in ["GET", "HEAD"] then
    allow
reject
\end{lstlisting}
\label{alg:RIP} 
\end{minipage}
\end{center}

\begin{center}
\begin{minipage}{0.95\linewidth}
\begin{lstlisting}[caption={Framing Isolation Policy (FIP).}, label={lst:pseudocode-fip}]
Input: req := HTTP request
if req.headers.secFetchMode != "navigate" then
    allow
if req.headers.secFetchDest not in ["frame", 
   "iframe", "embed", "object"] then
    allow
reject
\end{lstlisting}
\label{alg:FIP}
\end{minipage}
\end{center}

\begin{center}
\begin{minipage}{0.95\linewidth}
\begin{lstlisting}[caption={Strict Isolation Policy (SIP).}, label={lst:pseudocode-sip}]
Input: req := HTTP request
if req.headers.secFetchSite = "cross-site" then
    reject
allow
\end{lstlisting}
\label{alg:SIP}
\end{minipage}
\end{center}

\section{Custom Fetch Metadata Policies}
The following are pseudocode representations of custom Fetch Metadata policies identified during our measurement of the top 1M websites, as discussed in Sec.~\ref{subsec:resultsCustomPolicies}.

\begin{center}
\begin{minipage}{0.95\linewidth}
\begin{lstlisting}[caption={Custom Policy CP1.}, label={alg:document-frame-iframe}]
Input: req = HTTP request
if req.method = "POST" then
    allow
if req.headers.secFetchDest in ["document", 
   "frame", "iframe"] then
    allow
reject
\end{lstlisting}
\end{minipage}
\end{center}

\begin{center}
\begin{minipage}{0.95\linewidth}
\begin{lstlisting}[caption={Custom Policy CP2.}, label={alg:cross-site_none}]
Input: req = HTTP request
if req.headers.secFetchSite in ["cross-site", 
   "none"] and req.method = "POST" then
    reject
allow
\end{lstlisting}
\end{minipage}
\end{center}

\begin{center}
\begin{minipage}{0.95\linewidth}
\begin{lstlisting}[caption={Custom Policy CP3.}, label={alg:cross-site_object_embed}]
Input: req = HTTP request
if req.headers.secFetchSite = "cross-site" 
   and req.headers.secFetchDest in ["object", 
   "embed"] then
    reject
if req.method = "POST" and
   req.headers.secFetchSite = "cross-site" and
   req.headers.secFetchMode != "navigate" then
    reject
allow
\end{lstlisting}
\end{minipage}
\end{center}

\begin{center}
\begin{minipage}{0.95\linewidth}
\begin{lstlisting}[caption={Custom Policy CP4.}, label={alg:image}]
Input: req = HTTP request
if req.headers.secFetchDest in ["image"] then
    reject
allow
\end{lstlisting}
\end{minipage}
\end{center}

\section{Attacks}
\label{appendix:poc}
This appendix provides implementations of the timing attacks described in Sec.~\ref{chap:fetchComparison}. Listing~\ref{poc:iframe} shows an iframe-based attack which can be used to infer that state-dependent server-side computation occurred and is effective even when defenses like XFO, CSP and XCTO are enabled. Listing~\ref{poc:window} shows a variation of this attack using a window instead of an iframe, which is effective even when COOP is deployed.

\lstdefinelanguage{JavaScript}{
  keywords={typeof, new, true, false, catch, function, return, null, catch, switch, var, if, in, while, do, else, case, break, try, let},
  keywordstyle=\color{blue}\bfseries,
  ndkeywords={class, export, boolean, throw, implements, import, this},
  ndkeywordstyle=\color{darkgray}\bfseries,
  identifierstyle=\color{black},
  sensitive=false,
  comment=[l]{//},
  morecomment=[s]{/*}{*/},
  commentstyle=\color{purple}\ttfamily,
  stringstyle=\color{red}\ttfamily,
  morestring=[b]',
  morestring=[b]",
  morestring=[b]`
}

\lstset{
    language=javascript,
    basicstyle=\smaller\ttfamily,
    numbers=left,
    numberstyle=\tiny\color{gray},
    stepnumber=1,
    numbersep=5pt,
    frame=single,
    captionpos=b
}

\vspace{10pt}

\begin{minipage}{0.9\columnwidth}
\begin{lstlisting}[caption={XFO, CSP and XCTO PoC - Iframe-based XS-Leak.}, label={poc:iframe}]
<!DOCTYPE html>
<body><script>
function measure(url){
  var f=document.createElement('iframe');
  f.src=`${url}`;

  f.onload = () => {
    var time=performance.now()-start;
    console.log(`[${url}] ${time}$ms`);
  }
  document.body.appendChild(f);
  var start=performance.now();
};

window.onload = () => {
  measure("<xfo-protected-url>");
  measure("<csp-protected-url>");
  measure("<xcto-protected-url>");
};
</script></body>
\end{lstlisting}
\end{minipage}

\begin{center}
\begin{minipage}{0.9\columnwidth}
\begin{lstlisting}[caption={COOP PoC - Window-based XS-Leak.}, label={poc:window}]
<!DOCTYPE html>
<body><script>
  let coopUrl="<target-url>";
  var w=window.open(coopUrl, "_blank");
  var start=performance.now();
  function measure(){
    try{
      w.origin;
      setTimeout(measure, 0);
    }catch(e){
      var time=performance.now() - start;
      console.log(`${time} ms`);
   }
 }
 measure();
</script></body>
\end{lstlisting}
\end{minipage}
\end{center}

\section{Browser Constraints}
\label{appendix:browser-constraints}

This appendix specifies the complete set of constraints that define the browser-emittable set $\mathcal{R}_B \subseteq \mathcal{R}$. A request $r \in \mathcal{R}$ is browser-emittable ($r \in \mathcal{R}_B$) if and only if it satisfies all constraints below. These constraints are derived from the Fetch Standard~\cite{whatwg_fetch}, the HTML Standard~\cite{whatwg_html}, and the Fetch Metadata specification~\cite{west_20_fetch}.

\subsection{Navigation Constraints ($\mu = \text{navigate}$)}
Navigational requests have destination \code{document}, \code{embed}, \code{frame}, \code{iframe}, \code{fencedframe}, or \code{object}~\cite{whatwg_fetch}, and mode $\mu = \text{navigate}$.

\begin{enumerate}
    \item \textbf{Navigation Methods.} Browsers perform navigations via user actions (address bar, bookmarks, link clicks), HTML form submissions, or script APIs such as \code{window.location}, \code{location.assign}/\code{replace}, \code{form.submit()}, and \code{window.open}~\cite{whatwg_html}. All produce navigational requests ($\mu = \text{navigate}$); HTML restricts form submissions to \code{GET} and \code{POST}, and every other navigation trigger is uses \code{GET}. Script- and document-initiated navigations carry the initiator's origin, so $r.\sigma \in \{\text{same-origin}, \text{same-site}, \text{cross-site}\}$ and never $\text{none}$; only user-initiated navigations yield $r.\sigma = \text{none}$~\cite{west_20_fetch}.
    \begin{equation*}
    \forall r \in \mathcal{R}_B:\ (r.\mu = \text{navigate}) \implies r.m \in \{\text{GET}, \text{POST}\}
    \end{equation*}

    \item \textbf{User-Originated Requests.} When \code{Sec-Fetch-Site: none}, the request was directly initiated by the user (e.g., by typing an address or clicking a bookmark)~\cite{west_20_fetch}. Such navigations target the top-level browsing context, so $\delta = \text{document}$, and they are \code{GET}-only.
    \begin{equation*}
    \begin{aligned}
    &\forall r \in \mathcal{R}_B:\ (r.\sigma = \text{none}) \\
    &\quad \implies (r.\mu = \text{navigate} \land r.\delta = \text{document} \land r.m = \text{GET})
    \end{aligned}
    \end{equation*}

    \item \textbf{Plugin Resources.} \code{<object>} and \code{<embed>} cause navigation requests ($\mu = \text{navigate}$). Their \code{data}/\code{src} fetches are \code{GET}~\cite{whatwg_html}. \code{<object>} exposes a \code{name} attribute naming its content navigable, so a form can \code{POST} into it; \code{<embed>} has no \code{name}, cannot be a form target, and is \code{GET}-only:
    \begin{equation*}
    \begin{aligned}
    &\forall r \in \mathcal{R}_B:\ r.\delta = \text{embed} \\
    &\quad \implies (r.m = \text{GET} \land r.\mu = \text{navigate})
    \end{aligned}
    \end{equation*}

    \item \textbf{The CSRF Boundary.} A cross-site \code{POST} is emitted only as a form navigation, which may target a nested browsing context (e.g., a frame), or as a programmatic request~\cite{whatwg_html, whatwg_fetch}.
    \begin{equation*}
    \begin{aligned}
    &\forall r \in \mathcal{R}_B:\ (r.\sigma = \text{cross-site} \land r.m = \text{POST}) \\
    &\quad \implies \big((r.\mu = \text{navigate} \land r.\delta \in \{\text{document}, \text{frame}, \\
    &\qquad\qquad\qquad\text{iframe}, \text{object}\}) \\
    &\qquad\quad \lor\ (r.\delta = \text{empty})\big)
    \end{aligned}
    \end{equation*}
    The HTML form \code{target} attribute can direct a submission to any navigable container (\code{document}, \code{frame}, \code{iframe}, or \code{object}). \code{<embed>} cannot be a form target and is \code{GET}-only, so it is excluded. Programmatic cross-site \code{POST} ($\delta = \text{empty}$) is covered by the programmatic constraints below.
\end{enumerate}

\subsection{Subresource Constraints}
HTML elements that embed media, execute code, or load fonts trigger \emph{subresource} requests~\cite{whatwg_html}.

\begin{enumerate}
    \setcounter{enumi}{4}
    \item \textbf{Passive Subresources are GET-only.} The destinations \code{image}, \code{script}, \code{style}, \code{font}, \code{audio}, \code{video}, and \code{track} are fetched with \code{GET}. The default mode is \code{no-cors}. The mode is \code{cors} when the \code{crossorigin} attribute is present~\cite{whatwg_html}, and always \code{cors} for fonts~\cite{css_fonts_4}.
    \begin{equation*}
    \begin{aligned}
    &\forall r \in \mathcal{R}_B:\ (r.\delta \in \{\text{image}, \text{script}, \text{style}, \text{font}, \\
    &\qquad\qquad\quad \text{audio}, \text{video}, \text{track}\}) \\
    &\quad \implies (r.m = \text{GET} \land r.\mu \in \{\text{no-cors}, \text{cors}\})
    \end{aligned}
    \end{equation*}
\end{enumerate}

\subsection{Programmatic Constraints ($\delta = \text{empty}$)}
JavaScript APIs such as \code{fetch()} and \code{XMLHttpRequest}, use destination \code{empty}~\cite{whatwg_fetch}.

\begin{enumerate}
    \setcounter{enumi}{5}
    \item \textbf{Programmatic Requests.} Requests with destination \code{empty} use one of the programmatic modes, including WebSocket handshakes, which use mode \code{websocket}~\cite{rfc6455}. (WebTransport uses a Fetch mode that \code{Sec-Fetch-Mode} does not recognize, so it is excluded.)
    \begin{equation*}
    \begin{aligned}
    &\forall r \in \mathcal{R}_B:\ (r.\delta = \text{empty}) \\
    &\quad \implies r.\mu \in \{\text{cors}, \text{no-cors}, \text{same-origin}, \text{websocket}\}
    \end{aligned}
    \end{equation*}

\end{enumerate}

\end{document}